\documentclass[lettersize,journal]{IEEEtran}
\usepackage{amsmath,amsfonts}
\usepackage{algorithmic}
\usepackage{algorithm2e}
\usepackage{color}
\usepackage{tikz}
\usetikzlibrary{shapes.geometric, arrows, fit, positioning, calc, patterns}
\usepackage{array}
\usepackage{amsmath,amssymb}
\RestyleAlgo{ruled}
\usepackage{textcomp}
\usepackage{stfloats}
\usepackage{url}
\usepackage{verbatim}
\usepackage{graphicx}
\usepackage{cite}
\usepackage{xcolor}
\usepackage{float, flushend}
\usepackage{subcaption}
\definecolor{amethyst}{rgb}{0.6, 0.4, 0.8}

\begin{document}

\title{Active Sensing for RIS-Aided Tracking and Power Control:\\ A Hybrid Neuroevolution and Supervised Learning Approach
}

\author{George Stamatelis,~\IEEEmembership{Student~Member,~IEEE},  Hui Chen,~\IEEEmembership{Member,~IEEE}, Henk Wymeersch,~\IEEEmembership{Fellow,~IEEE},\\ and George C. Alexandropoulos,~\IEEEmembership{Senior~Member,~IEEE}
\thanks{
A preliminary version of this manuscript has been presented at the IEEE ICASSP, Barcelona, Spain, May 2026~\cite{ConfVersionPowerControl}.
}
\thanks{G. Stamatelis and G. C. Alexandropoulos are with the Department of Informatics and Telecommunications, National and Kapodistrian University of Athens, Panepistimiopolis Ilissia, 16122 Athens, Greece (e-mails: \{georgestamat, alexandg\}@di.uoa.gr).} 
\thanks{H. Chen and H. Wymeersch are with the Department of Electrical Engineering, Chalmers University of Technology,
412 58 Gothenburg, Sweden (e-mails: \{hui.chen, henkw\}@chalmers.se).}
\thanks{This work has been supported by the Smart Networks and Services Joint Undertaking project 6G-DISAC under the European Union's Horizon Europe research and innovation programme under Grant Agreement No 101139130. G. Stamatelis was also supported by the Hellenic Foundation for Research and Innovation (HFRI) under the 5th Call for HFRI PhD Fellowships (Fellowship Number: 21080).
}
}



\maketitle

\begin{abstract}
This paper studies energy efficient tracking of power-limited mobile users with the assistance of a Reconfigurable Intelligent Surface (RIS). Since localization pilot transmissions dominate the energy budget of power-constrained devices, we introduce a low-overhead feedback link from the Base Station (BS) to the user to enable dynamic uplink power control. To navigate the discrete and decentralized nature of this active sensing problem, we propose a novel Dual-Agent (DA) deep learning framework that jointly optimizes the discrete RIS phase profiles and the UE's transmit power in real time. Specifically, our approach employs a hybrid training methodology integrating the neuroevolution paradigm with supervised learning, effectively overcoming the non-differentiability of discrete phase responses from the RIS unit elements and the strict information bottleneck of single-bit feedback messages for pilot power control. The proposed DA active sensing framework can be applied with both single- and multi-antenna BSs, the latter with only minor modifications in the structure of one NN: an additional output branch with appropriate structure is included for the latter case to select a valid digital combiner from a finite set. Extensive numerical simulations demonstrate that the proposed scheme achieves highly accurate and robust tracking across diverse target motion models, outperforming extended Kalman and particle filters, as well as, machine learning-based trackers. Furthermore, in static localization, it is shown to significantly outperform traditional fingerprinting schemes, deep reinforcement learning baselines, and standard backpropagation-based estimators.
\end{abstract}

\begin{IEEEkeywords}
Tracking, beamforming, power control, multi-agent systems, neuroevolution, reconfigurable intelligent surface.
\end{IEEEkeywords}

\section{Introduction}
Precise device and object location knowledge can significantly enhance modern wireless applications, including robotic navigation \cite{localizationRobotics}, intelligent vehicles and traffic management \cite{localizationVehicles,localizationTraffic-AV}, Internet of Things (IoT) \cite{IoTLocalization1,IoTLocalization2}, as well as assisted living \cite{localizationAssistedLiving}. However, accurate location estimation in multipath environments is a challenging problem, especially when trying to localize power-limited lightweight IoT devices. To this end, Reconfigurable Intelligent Surfaces (RISs)~\cite{10596064} have emerged as a powerful candidate for enhancing the performance of various communication, localization, and sensing schemes~\cite{RISCRB-loc,ris_passive_loc_ref1,ris_passive_loc_ref2,fingerprint,LeveragingRisVehicular,RIS_sidelink}.

An RIS consists of a large number of low-cost, passive elements  that can intelligently manipulate electromagnetic waves, typically radio waves, to enhance the performance of wireless networks. When a signal is transmitted, it encounters obstacles or interference, and the RIS intelligently manipulates the reflection of these signals to enhance their propagation. By controlling the characteristics of the reflected waves, RIS can direct the signal to specific locations, improve coverage, reduce interference, and enhance overall signal quality. This adaptive manipulation of signals is achieved with low-power consumption, making RIS an efficient solution for optimizing wireless communication in complex environments, such as urban areas or indoor spaces. Naturally, RISs are envisioned as one of the core technologies for the upcoming (6G) generation of wireless networks\cite{Huang_Reconfigurable_2019,RIS_arch2,towardsIRS}.

However, in order to harvest the full potential of this technology, careful tuning of their elements is necessary. Most practical hardware RIS implementations come with discrete and even binary RIS phase profiles~\cite{RIS_SRE,1Bit_Sign_Alignment}, making it practically impossible to find optimal solutions, as discrete optimization is NP-hard. To that end, various approximate solution methods for a wide variety of applications have been proposed, most of whom are based on machine learning. For instance, Deep Reinforcement Learning(DRL) \cite{RIS_DRL1,Stylianop1Bit,AlexandroPervasive}, neural contextual bandits\cite{stylianop-MAB-ICC}, and even NeuroEvolution (NE)\cite{mbacnnJournal} have been applied to adaptive RIS control problems.

\subsection{Background and Related Works}
\paragraph{Active Sensing and Localization}
Active sensing refers to the adaptive, online reconfiguration of environmental parameters to enhance the performance of a specific estimation task. The theoretical foundations of this field were established by Chernoff in his seminal work on Active Hypothesis Testing (AHT)\cite{ChernoffHT}. Chernoff introduced an information-theoretic criterion for sequentially selecting the most informative experiments within a hypothesis testing framework. While his original formulation focused on optimizing medical trials, the underlying model has since found broad applicability in wireless communications, including radar-assisted target classification \cite{ChernoffRadar}, and sensor networks \cite{CohenAHT_AD}. More recently, research in signal processing has successfully bridged classical theory with modern learning-based approaches, combining traditional AHT formulations with data-driven artificial Neural Network (NN) algorithms (see, for example,~\cite{active_sensing_ref1,active_sensing_ref2,active_sensing_ref3,eaht}.

In position estimation problems, like the ones considered in this paper, a single User Equipment (UE) transmits pilot signals to a Base Station (BS), and the latter receives the positioning pilot signals through their direct link and reflections from scatterers. To this end, RISs constitute an important infrastructure component in environments where multiple obstacles obstruct the direct UE-BS path~\cite{RIS_SRE}, enabling dynamically programmable RIS-induced reflections. In such RIS-empowered systems, active sensing indicates adaptive online selection of the elements of the RIS. In fact, it has been experimentally shown to significantly outperform passive RIS-enabled localization algorithms \cite{activeSensingLocJournal}. 

More specifically, \cite{activeSensingLocJournal} designed an NN algorithm for jointly manipulating the RIS profiles actively (control task), and estimating the true UE location (estimation task). The algorithm, based on Long Short Term Memory (LSTM) models \cite{LSTM}, maintains a hidden state  representing important knowledge from past observation, and uses that along with the most recent observation in order to select the next RIS phase profile. Once transmissions have stopped, hidden state features are passed to another NN  responsible for estimating the UE's position. The entire parameter set (RIS selection module and estimation module) were trained as one network with backpropagation based optimizers. 
An extension of this algorithm for localization services with privacy risks was proposed, very recently, in \cite{PrivateLocICASSP}, where the RIS was tasked to avoid sending strong reflections to problematic areas where malicious third parties are located. A Lagrangian-based optimization algorithm that balances an estimation objective with a privacy criterion was designed in order to enforce location leakage constraints.

However, the aforedescribed research works are limited to idealized RIS models ignoring practical hardware impairments and limitations. In practical hardware models, RIS elements are typically limited to finite sets of phase configuration values due to hardware quantization~\cite{RIS_arch1,RIS_arch2,1Bit_Sign_Alignment,HW_generic1,HW_generic2,RIS_SRE}, which introduces additional challenges. To this end, at each time instance, the NN controlling the RIS has to output a discrete-valued vector corresponding to the next phase profile, and future profile selections, as well as the final estimation depend on these outputs, meaning that the optimization objective becomes non-discrete (and hence non-differentiable). This feature implies that typical backpropagation optimizers cannot be applied to such objective functions. In addition to overlooking these discrete hardware constraints, existing approaches are strictly limited to static localization scenarios. To bridge this gap, in this paper, we present a novel estimation framework that extends beyond stationary positioning to enable the real-time tracking of mobile trajectories in multipath environments.

The state-of-the-art assumes fixed transmission power for the localization pilot signals, which can be highly inefficient for dynamic trajectories. In practical IoT deployments, devices operate under strict energy constraints where continuous, high-power pilot transmissions rapidly drain battery life~\cite{AdHocNetw_EnergyConserv,SPM_data_transmission_expensive,EnergyCostThousand}. Conversely, a fixed low-power transmission risks losing track of a UE of interest during sudden maneuvers or severe fading. To resolve this, we herein introduce dynamic power control into the active sensing loop. By leveraging the LSTM's hidden state, which inherently encapsulates the system's uncertainty regarding UE's trajectory, our framework is designed to learn to adjust pilot transmission power intelligently over time. In particular, it conserves energy during predictable movement phases, and selectively boosts transmission power only when necessary to refine the location estimate or recover from severe attenuation. This joint optimization of discrete RIS configurations and temporal power allocation establishes a highly energy-efficient tracking protocol, perfectly aligned with the resource-constrained requirements of 6G IoT networks.

\paragraph{RIS-Assisted Tracking Algorithms}
Besides static UE location estimation, RISs have been also used to improve the performance of tracking systems due to their inherent ability to dynamically improve coverage on different areas~\cite{JSTP_RIS_Tracking,ElsevierRISTracking,MDPI_RIS_Tracking}. However, the reliance on unrealistic RIS response models remains a critical bottleneck in the existing tracking literature: state-of-the-art algorithms predominantly assume continuously tunable phase responses to facilitate real-time optimization. Beyond these hardware constraints, current works typically demand computationally prohibitive operations at every time step (such as message passing on graphs, heavy matrix inversions, and iterative gradient descent), further hindering their practical applicability. Particularly, in scenarios with high UE mobility, this computational latency is unacceptable, since a fast-moving target will have already transitioned to a new state by the time the algorithm converges. Finally, the available methods are strictly tailored to fully specified analytical channel models, limiting their generalizability to complex, real-world propagation environments. Conversely, data-driven approaches offer a robust alternative; they are inherently model-agnostic and capable to learn the underlying propagation mechanics autonomously, completely bypassing the need to hardcode rigid analytical channel expressions into the NN architecture~\cite{activeSensingLocJournal}.

\color{black}
\paragraph{Learning-Based System Configuration}
NE algorithms reframe the training of NN models as a stochastic global search problem, which is typically solved via evolutionary or genetic strategies. These methods maintain a population of individuals, each representing the parameters of a policy NN, and iteratively refine this population using evolution-inspired operators~\cite{introtoEA}. While the application of NE to sequential decision-making is a well-established concept~\cite{moriarty:mlj96, deepPOMDPsSchmidhuber}, it has recently been demonstrated that this tool can rival, or even surpass, state-of-the-art DRL algorithms~\cite{such2018deep,salimans2017evolution}, establishing a robust benchmark in wireless communication domains~\cite{eaht,mbacnnJournal}. Notably, even very old and simple evolutionary algorithms can yield near state-of-the-art performance on popular DRL testbeds~\cite{benchmaringBasics}.

A paramount advantage of NE is its gradient-free nature, which avoids the instability issues—such as vanishing or exploding gradients—that often plague typical deep learning and DRL optimization schemes associated with backpropagation through time~\cite{deepPOMDPsSchmidhuber}. This characteristic is particularly advantageous for online RIS configuration control, given the large number of the constituent metamaterials with quantized electromagnetic responses; such discrete constraints create a rugged optimization landscape that is not naturally compatible with differentiation-based algorithms. Furthermore, although DRL can provide strong decision-making capabilities, its effectiveness often hinges on the availability of well-defined, dense reward functions. Solving problems with multiple conflicting constraints makes the construction of such reward signals challenging and limiting. In contrast, NE operates effectively with sparse feedback, requiring only a scalar fitness score to be evaluated at an episode's conclusion.

\subsection{Paper's Contribution}
The contributions of this paper are summarized as follows:\begin{itemize}
    \item  \textbf{Problem definition:} We formulate the active uplink tracking problem for power-limited UEs aided by an RIS. Distinct from previous studies, our model strictly incorporates practical hardware and system constraints; in particular, \textit{i}) discrete phase shifts for the RIS elements, and \textit{ii}) a limited feedback capacity for uplink power control, restricting the BS-UE control link to messages of\footnote{As will be explained later on, the proposed uplink power control frameowrk requires only trivial modifications to incorporate multi-bit feedback. However, we chose to focus on the more challenging case of single-bit feedback accommodating to ultra-low-power IoT receivers~\cite{UltraLowPoer,WakeUpReceivers}.
    } a single bit. Besides localization of static UEs considered by prior works~\cite{activeSensingLocJournal,PrivateLocICASSP} we also consider position estimation for moving UEs.
    \item \textbf{Novel multi-agent algorithm:} We develop a novel Dual-Agent (DA) NE framework to solve the proposed non-differentiable joint control problem of the RIS phase configuration and the uplink UE transmission power. The algorithm features two collaborating Recurrent NNs (RNNs): \textit{i}) a BS agent that actively tunes the discrete RIS configuration, and \textit{ii}) a UE agent that adaptively manages pilots transmit power by interpreting the history of binary feedback. By optimizing these agents via evolutionary strategies, we bypass the non-differentiability issue of discrete hardware, enabling efficient joint learning.
\end{itemize}
Extensive numerical experiments are presented revealing that the proposed algorithm strictly satisfies the required power budget constraints, while achieving superior performance compared to traditional filters~\cite{EKF-ref1,PF1} and supervised RNN trackers. Furthermore, on static localization, our approach outperforms fingerprinting~\cite{fingerprint}, purely supervised, and DRL baselines. In addition, our simulation studies demonstrate that our single-bit feedback algorithm virtually matches the performance of NE variants utilizing expensive scalar control links, and exhibits minimal performance drop-off compared to  schemes using full transmission power for the pilot signals.

Compared to the preliminary conference version in~\cite{ConfVersionPowerControl}, this work extends the power-constrained active sensing framework to support the continuous tracking of mobile UEs and the configuration of multi-antenna BSs. In addition, a comprehensive performance evaluation under varying scattering conditions, including a sensitivity analysis of the NE hyperparameters, is presented. Furthermore, we investigate the design trade-offs of the single-bit control link, demonstrating how the proposed learned collaborative protocol mitigates the information bottleneck to experimentally outperform traditional tracking filters \cite{EKF-ref1,PF1} as well as localization baselines~\cite{fingerprint}.

\subsection{Notation and Organization}
Lower case bold letters are used to represent vectors, e.g. $\mathbf{x}$, and upper case bold letters are reserved for matrices, e.g. $\mathbf{X}$. The 
conjugate transpose of a matrix $\mathbf{X}$ is denoted as $\mathbf{X}^{\rm H}$, whereas $\text{diag}(\mathbf{x})$ denotes the diagonal matrix constructed by the vector $\mathbf{x}$. $\mathbf{I}_d$ represents the $d$-dimensional identity matrix and $\mathbf{0}_d$ a $d$-dimensional vector filled with zeros. Calligraphic letters, e.g. $\mathcal{X}$, typically represent sets, and $\boldsymbol{w}_x$ represents the trainable weights of a NN, stacked in a vector. $
\mathbb{E}[\cdot]$ denotes expectation, whereas $\Re(x)$ and $\Im(x)$ represent the real and imaginary parts of a complex number $x$, respectively. Finally, and the cardinality of a set $\mathcal{X}$ is denoted as $\rm{card}(\mathcal{X})$.

The rest of this paper is organized as follows. Section \ref{sec:systemmodel} presents the wireless system under consideration and the power-constrained UE tracking problem. The proposed multi-agent algorithm is described in Section~\ref{sec:multiDL} and numerically verified in Section~\ref{sec:examples}. Finally, Section~\ref{sec:conclusion} concludes the paper.

\section{System Modeling and Design Objective}\label{sec:systemmodel}
We consider a system comprising a single-antenna\footnote{The single-antenna BS case has been chosen, herein, for ease of exposition and in order to be consistent with the system models of relevant prior works~\cite{activeSensingLocJournal,PrivateLocICASSP}. Later on, in Section~\ref{sec:multi_antenna_extension}, the system model and problem formulation will be extended to the multi-antenna BS case.} BS receiving pilot symbols from a mobile single-antenna UE positioned in an unknown location $\mathbf{p}_t\in \mathbb{R}^3$, with the intention to obtain an accurate estimate for $\mathbf{p}_t$, denoted, henceforth, as $\hat{\mathbf{p}}_t$, at each time slot $t$. In particular, the UE is assumed to move according to a prespecified model $m(\cdot)$, as follows: \begin{equation}
    \label{eq:target_state_update}
    \mathbf{x}(t) \triangleq m(\mathbf{x}({t-1}))+\bar{\mathbf{n}}(t) \in \mathbb{R}^s,
\end{equation}
where the $s$-dimensional state vector $\mathbf{x}(t)$ collects the target coordinates $\mathbf{p}_t$ as well as other relevant target information, such as velocities, turn rate, and acceleration. The noise vector $\bar{\mathbf{n}}(t)$ is typically assumed to be zero-mean Gaussian, i.e., $\bar{\mathbf{n}}(t) \sim \mathcal{N}(\mathbf{0}_s,\sigma^2_{\rm ue} \mathbf{I}_s)$, and accounts for modeling, estimation, and generalization errors. We finally use notation $m_{\rm prior}$ to represent any state prior distribution available. 
\begin{figure}[t]
    \centering
    \includegraphics[width=\linewidth]{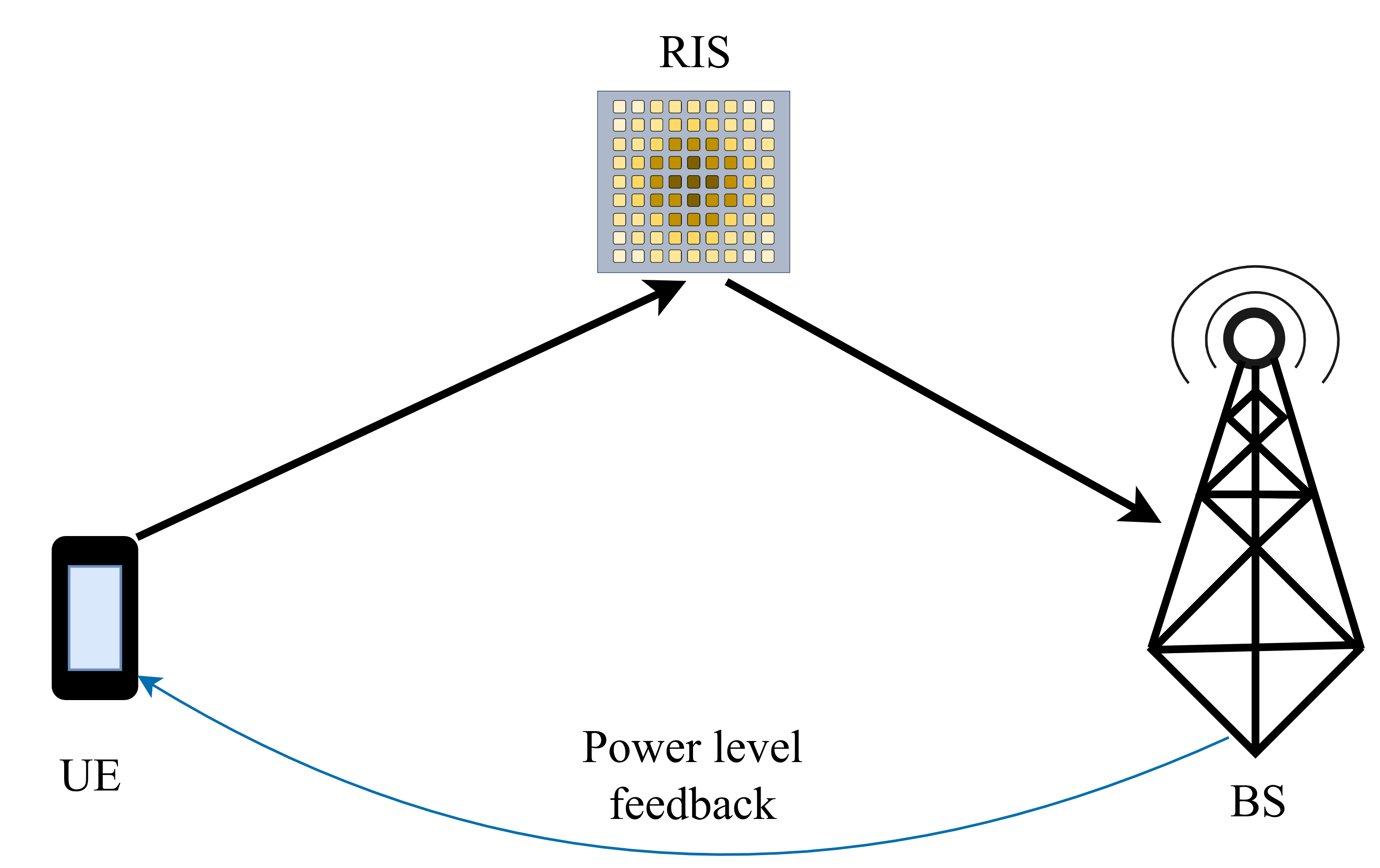}
    \caption{The considered RIS-aided active UE tracking system incorporating pilot transmission power control.}
    \label{fig:placeholder}
\end{figure}

The pilot symbols communication is assisted by an RIS whose controller is managed by the BS via a dedicated error-free control channel~\cite{ControlSignaling}. The RIS consists of $N_{\rm ris}$ response-tunable elements, which, according to the vast majority of the currently available hardware implementations \cite{RIS_arch1,RIS_arch2,1Bit_Sign_Alignment,HW_generic1,HW_generic2,RIS_SRE}, contribute an effective discrete phase shift on the impinging signal. We denote this phase configuration for each $i$-th RIS unit element ($i=1,\ldots,N_{\rm ris}$) as $\theta_i$, with all belonging to a finite set $\Theta$. The static positions $\mathbf{p}_{\rm bs}$ and $\mathbf{p}_{\rm ris}$ of the BS and the RIS, respectively, are assumed to be known to the estimation framework presented in this paper. As the UE follows the dynamic motion model in~\eqref{eq:target_state_update}, the wireless channels are re-sampled at each time instance $t$, independently reflecting the evolving geometry of the BS-UE, BS-RIS, and RIS-UE wireless links. The UE transmits a sequence of $T$ pilot symbols to the BS over an equal number of consecutive time instances. At each $t$-th frame ($t=1,\ldots,T$), both the UE's transmission power $P(t)$ and the RIS phase configuration profile, denoted as $\boldsymbol{\Phi}(t)\triangleq\text{diag}(\boldsymbol{\phi}(t))$ with:
\begin{equation}
    \boldsymbol{\phi}(t)\triangleq[e^{j\pi\theta_1(t)},\ldots,e^{j\pi\theta_{N_{\rm ris}}(t)}],
\end{equation}
constitute free design parameters that we intend to  optimize for our UE tracking objective.

\subsection{Received Signal Model}
Let $x(t)$ denote the unit-power complex-valued pilot symbol transmitted at each $t$-th time instance. The baseband received signal at the BS side during this frame is  modeled as follows:
\begin{equation}
    \label{eq:received_signal}
    y(t)\!\triangleq\!\sqrt{P(t)} \left(h_d(\mathbf{p}_t)\!+\!\mathbf{h}_{\rm{bs},\rm{ris}}(\mathbf{p}_t) \boldsymbol{\Phi}(t) \mathbf{h}_{\rm{ris},\rm{ue}}(\mathbf{p}_t)\right)\!x(t)+n(t),
\end{equation}
where $h_d (\mathbf{p}_t) \in\mathbb{C}$ is the gain of the direct BS-UE channel, whereas $\mathbf{h}_{\rm{bs},\rm{ris}}(\mathbf{p}_t)\in\mathbb{C}^{1\times N_{\rm ris}}$ and $\mathbf{h}_{\rm{ris},\rm{ue}}(\mathbf{p}_t)\in\mathbb{C}^{N_{\rm ris}\times1}$ represent the respective gains of the BS-RIS and RIS-UE channel matrices, and $n(t)\sim\mathcal{N}(0,\sigma^2_n)$ is the Additive White Gaussian Noise (AWGN), whose variance $\sigma^2_{n}$ can be reliably estimated, thus, assumed known. To simplify the notation, in the remainder of the paper, we will drop the term $\mathbf{p}_t$ from the symbols for the channels and related measures, implying that all of them depend on the unknown UE position.

\subsection{Channel Model}
We adopt a general Ricean fading model for all wireless channels, which accounts for the presence of a dominant Line-of-Sight (LoS) component alongside Non-LoS (NLoS) multipath scattering~\cite{MatthaiouandGA_Fading}. Specifically, the channel vector (or scalar) $\mathbf{h} \in \{h_{d}, \mathbf{h}_{\rm bs,ris}, \mathbf{h}_{\rm ris,ue}\}$ is expressed as follows:
\begin{equation}
    \mathbf{h} = \sqrt{\beta}\left( \sqrt{\frac{\kappa}{\kappa+1}}\mathbf{h}^{\text{LoS}} + \sqrt{\frac{1}{\kappa+1}}\mathbf{h}^{\text{NLoS}} \right),
    \label{eq:ricean_model}
\end{equation}
where $\beta$ denotes the distance-dependent path loss dictated by the UE position at time $t$, $\kappa$ represents the Ricean factor, and $\mathbf{h}^{\text{LoS}}, \mathbf{h}^{\text{NLoS}}$ denote the deterministic Line of Sight (LoS) and stochastic Non-LoS (NLoS) components, respectively. The NLoS components are modeled as standard Rayleigh fading channels with entries drawn from a complex Gaussian distribution $\mathcal{CN}(0, 1)$. The LoS components are defined by the geometric array response vectors corresponding to the Angles of Departure (AoD) and Arrival (AoA), which are determined by the positions $\mathbf{p}_t, \mathbf{p}_{\rm bs},$ and $\mathbf{p}_{\rm ris}$. To account for the obstruction of the direct path,  the direct link $h_d$ is modeled with an additional blockage attenuation coefficient $\rho \in (0,1]$, such that $h_d = \sqrt{\rho}\,h_{d, \text{Ricean}}$; factor $\rho$ captures the penetration loss caused by obstacles (e.g., walls or trees) obscuring the LoS propagation path between the UE and the BS. This attenuation is introduced to simulate challenging tracking scenarios. Note that, in the presence of a strong LoS direct path, moderate-to-high received Signal-to-Noise Ratio (SNR) levels would typically suffice for accurate UE tracking\cite{AlexandroBackhaul}, rendering the active RIS configuration redundant.
\subsection{Proposed Active Sensing Protocol}
Upon collecting the received signal $y(t)$ at each $t$-th time instance, the BS decides on the RIS phase configuration for the next time instance~\cite{activeSensingLocJournal}, $\boldsymbol{\Phi}(t+1)$, as well as on the value of a control variable $b(t)$ to be fed back to the UE instructing it to refine its pilot symbols transmission power at the next time instance, $P(t+1)$. More specifically, when the received observations are informative enough for accurate tracking, the BS may request a lower uplink transmission power, thus,  enabling power savings at the UE side. On the other hand, when the received signals are too noisy to infer $\mathbf{p}_t$ satisfactorily, the BS requests higher power levels from the UE for its future pilot symbols transmissions. 

To facilitate robust detection of the lightweight feedback messages, we restrict $b(t)$ to single-bit commands (e.g., `0' implies power reduction and `1' power boosting), leaving the UE to dynamically determine its exact transmission power within a range $[0, P_{\max}]$. While multi-bit feedback could explicitly dictate precise power levels, decoding such complex messages typically requires coherent demodulation at the UE, which demands power-hungry phase tracking hardware. Alternatively, while non-coherent amplitude tracking allows for ultra-low-power reception at the UE side (e.g., wake-up or low-rate control receivers~\cite{WakeUpReceivers,UltraLowPoer}), attempting to decode multi-bit symbols non-coherently introduces a dimensionality constraint that may demand prohibitively large BS transmit power levels to maintain distinguishable thresholds~\cite{NonCoherentMimo}. Consequently, designing our collaborative protocol around $1$-bit feedback ensures the UE can utilize an ultra-low-power non-coherent receiver without imposing excessive transmit power requirements on the BS. To compensate for this strict information bottleneck, the proposed active sensing protocol relies on the UE to intelligently integrate the entire history of its received feedback bits. By interpreting the temporal sequence of binary commands rather than just the instantaneous bit, the UE may infer the urgency of the sensing task and adjust its power accordingly. As we will experimentally demonstrate, this proposed learned protocol incurs minimal performance degradation, allowing the proposed lightweight algorithm to virtually match the tracking accuracy of policies utilizing substantially costlier, high capacity control links.

\subsection{Problem Formulation}
According to the adopted active sensing paradigm, the BS decides at each $t$-th time instance the next frame's RIS phase configuration, $\boldsymbol{\Phi}(t+1)$, leading to the most favorable observations for the UE tracking objective. In addition, it also decides the transmit power level $P(t+1)$, which is acknowledged to the UE via the $b(t)$ transmission in the control/feedback channel. This mode of operation implies that the RIS phase profile and the UE transmit power levels at each $(t+1)$-th frame depend on all past observations. Let us define this dependency through a function $g(\cdot)$, i.e., it holds $\forall t<T$:
\begin{equation}
    \label{eq:policyDef}
    \left\{P(t+1),\boldsymbol{\Phi}(t+1)\right\}\triangleq g\left(y(1),\ldots,y(t)\right).
\end{equation}
The initializations $P(0)$ and $\boldsymbol{\Phi}(0)$ can be set to arbitrary values (e.g., $P(0)=P_{\rm max}$, and $\boldsymbol{\Phi}(0)$ so as to illuminate a large portion of the RIS area of influence~\cite{9827873}) if there is lack of any relevant a priori information for the tracking objective. Function $g(\cdot)$ that actually determines the active sensing policy can be separated in two mappings: the BS mapping $g_{\rm bs}(\cdot)$ in charge of configuring RIS profiles and selecting the $1$-bit feedback messages, and the UE mapping $g_{\rm ue}(\cdot)$. Formally: 
\begin{subequations}\label{eq:twoPolicyDef}
    \begin{align}
  \{\boldsymbol{\Phi}(t+1),b(t)\}&\triangleq  g_{\rm bs}(y(1),y(2),\ldots,y(t)), \label{eq:twoPolicyDef_a}\\
  P(t+1)&\triangleq g_{\rm ue}(b(1),b(2),\ldots,b(t)).\label{eq:twoPolicyDef_b}
\end{align}
\end{subequations}
Finally, following the same mindset, the UE position estimation at the BS after the reception of the $t$ pilot symbols will be a function of these symbols (i.e., the processing result upon them). Let $f(\cdot)$ represent this function, hence, we define: 
\begin{equation}\label{eq:decDEf}
    \hat{\mathbf{p}}_t\triangleq f\left(y(1),\ldots,y(t)\right).
\end{equation}

Let $\mathcal{G}_{\rm bs},\mathcal{G}_{\rm ue}$ denote, respectively, the set of all admissible functions $g_{\rm bs}(\cdot),g_{\rm ue}(\cdot)$ in \eqref{eq:twoPolicyDef}, and $\mathcal{F}$ the set of all estimator functions $f(\cdot)$ in \eqref{eq:decDEf}. Adopting the Euclidean distance metric (specifically, the Mean Squared Error (MSE)), we formulate the following optimization problem for our tracking objective:
\begin{subequations}
\begin{equation}\label{eq:objective}
 \mathcal{OP}:    \min_{g_{\rm bs},g_{\rm ue} ,f(\cdot)} \mathbb{E}\left[\sum_{t=1}^T \left\|\hat{\mathbf{p}}_t-\mathbf{p}_t\right\|_2^2\right]
\end{equation}
\begin{align}
  \text{s.t.} \quad   &\theta_i(t) \in \Theta \,\,\forall i=1,\ldots,N_{\rm ris}, \forall t=1,\ldots,T, \label{eq:risconstraint}\\ 
    &P(t) \in [0,P_{\rm max}] \,\,\forall t=1,\ldots,T, \label{eq:pmaxconstr}\\ 
    &E\left[\sum_{t=1}^T P(t)\right] \leq B_P, \label{eq:ptotalconstr}
\end{align}
\end{subequations}
where $B_P$ represents a cumulative UE power budget constraint over the considered $T$-frame time horizon. It is noted that an interesting special case of the $\mathcal{OP}$ formulation, that has received the most attention in the relevant active sensing literature~\cite{activeSensingLocJournal,PrivateLocICASSP}, is the ``static'' localization case. 
In this case, the estimator $f(\cdot)$ is only used in the final instance of the $T$-instance horizon, when all received signal information has been collected, and the most accurate possible estimate can be made. Denoting this estimate as $\hat{\mathbf{p}}=f(y(1),\ldots,y(T))$, $\mathcal{OP}$'s objective simplifies, in the ``static'' localization case, to:  \begin{equation}
    \label{eq:passive_objective}
    \min_{g_{\rm bs}(\cdot),g_{\rm ue}(\cdot), f(\cdot)} \mathbb{E}\left[ 
    \left\|\hat{\mathbf{p}}-\mathbf{p} \right\|_2^2
    \right],
\end{equation}
while the constraints \eqref{eq:risconstraint}--\eqref{eq:ptotalconstr} remain the same.

\section{The Proposed DA Deep Learning Method}\label{sec:multiDL}
\begin{figure}
    \centering
    \includegraphics[width=\linewidth]{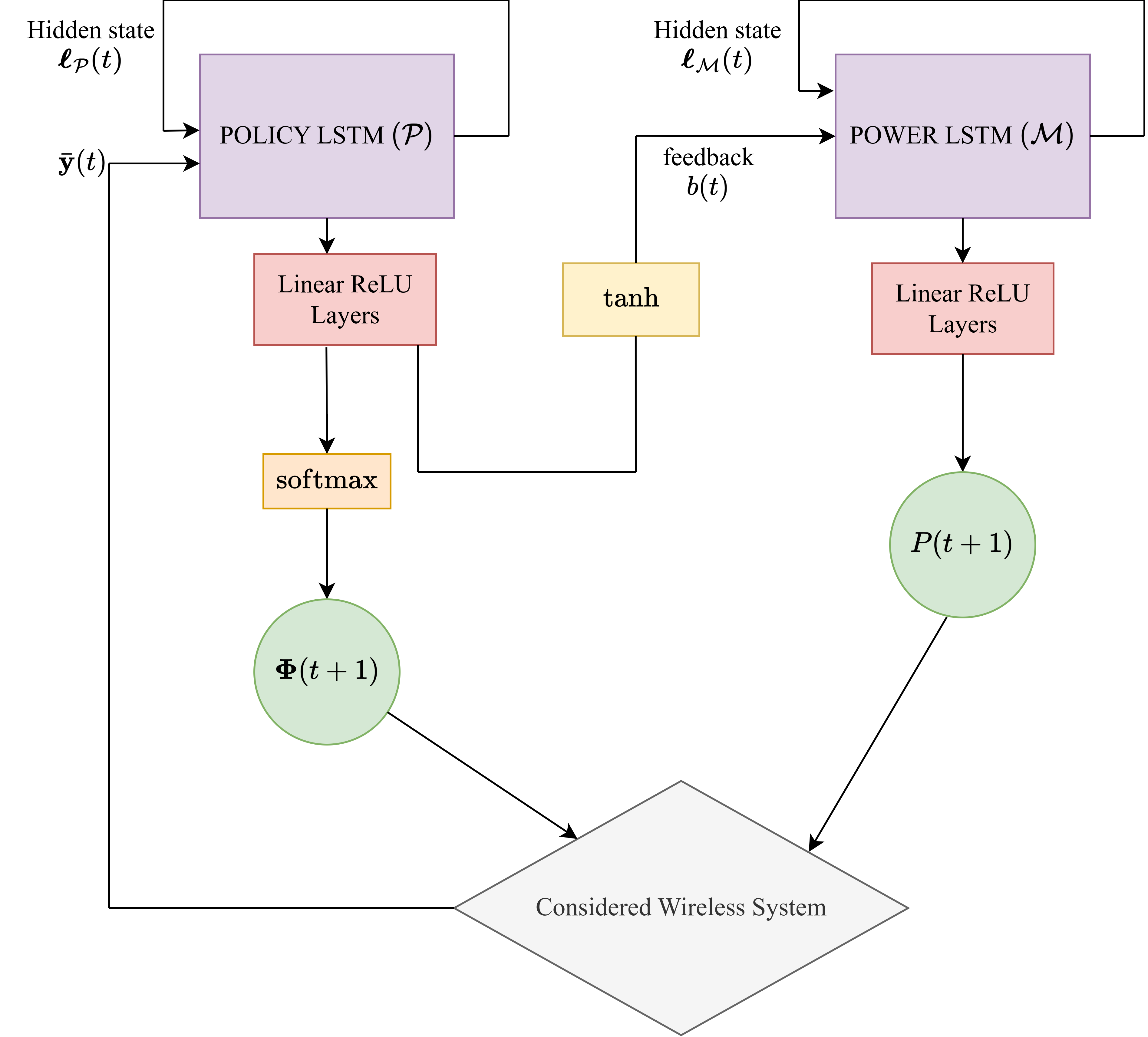}
		\caption{Graphical illustration of the  
 DA algorithm describing the collaborating agents at the BS and UE sides.}
    \label{fig:multiNet}
\end{figure}
In this section, we present our DA approach for the considered binary feedback messages enabling the UE power control in the uplink. As illustrated in Fig.~\ref{fig:multiNet} and detailed in the sequel, the proposed approach involves two collaborating agents: one at the BS side and one at the UE. Their policies are jointly optimized to achieve the sufficient level of coordination enabling accurate UE localization and tracking. 

\subsection{Data-Driven $\mathcal{OP}$ Solution}
Solving $\mathcal{OP}$ analytically is computationally prohibitive. At every time instance $t$, the optimal strategy requires an exhaustive search over the combinatorial space of discrete RIS phase configurations (which scales as ${\rm card}(\Theta)^{N_{\rm ris}}$), coupled with the continuous UE power allocation variable. This instantaneous optimization is further complicated by the following temporal dependency: current actions dictate the quality of future estimates, necessitating a dynamic programming approach rather than greedy selection. Formally, this falls under the umbrella of Partially Observable Markov Decision Processes (POMDPs)~\cite{krishnamurthy2025partially}. The partial observability arises particularly because the true system state, i.e., the unknown UE's position $\mathbf{p}_t$, is hidden and must be inferred from noisy observations (following the model in~\eqref{eq:received_signal}). Exact solutions to POMDPs generally require tracking a continuous, high-dimensional belief state (the posterior distribution of $\mathbf{p}_t$), a task proven to be NP-hard\cite{papadimitriou1987complexity} even without our additional UE power budget constraint.

In more detail, our UE tracking problem formulation essentially constitutes a Decentralized POMDP (Dec-POMDP), which introduces exponential complexity\cite{bernstein2002complexity}. Actually, this complexity stems from the asymmetric information structure: the BS observes $y(t)$, but cannot directly set the UE pilot transmission power, while the UE controls the power but observes only the single-bit feedback message $b(t)$. These two agents must, therefore, implicitly learn a coordination protocol to overcome this information bottleneck. To this end, inspired by prior works on related active sensing problems, e.g., \cite{active_sensing_ref1,active_sensing_ref2,active_sensing_ref3}, we devise a data-driven algorithm to approximate the intractable optimal policy. We leverage the ability of RNNs to maintain a compact hidden state that acts as a surrogate for the complex belief state. In the sequel, we parametrize the adaptive control policies $g_{\rm bs}(\cdot),g_{\rm ue}(\cdot)$ and the position estimator $f(\cdot)$ as Deep NNs (DNNs), and, to navigate the non-differentiable landscape due to discrete RIS phase responses, we propose a three-step training approach for their efficient training.

\subsubsection{The BS Agent} A \textit{policy} NN, $\mathcal{P}$
and an \textit{estimator} NN, $\mathcal{E}$ 
are deployed, with the former controlling the RIS phase configuration as well as the BS's binary feedback to the UE, and the latter being responsible for processing the observation sequence and estimating the UE location. Both NNs include LSTMs and, for simplicity, we assume that they share the same number of hidden layers, activation functions, and layer sizes. However, they differ in their subsequent Feed-Forward (FF) stacks, which are tailored to their output requirements. 

\begin{figure}
    \centering
    \includegraphics[width=\linewidth]{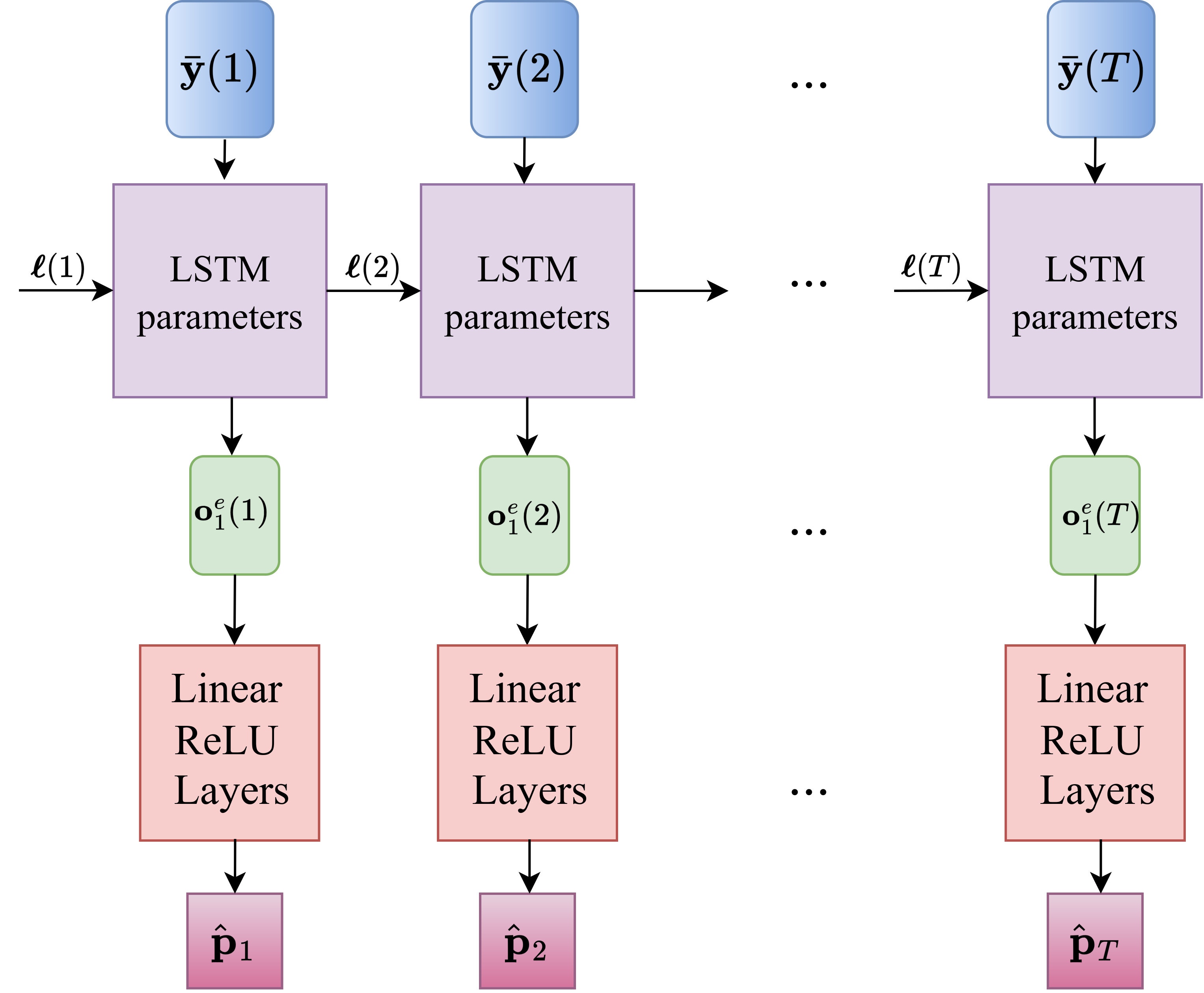}
    \caption{A high-level overview of the estimator NN $\mathcal{E}$ at the BS: At each time instance $t$, the vector $\bar{\mathbf{y}}(t)$ is passed through the LSTM parameters, and the resulting hidden state is stored for the next time instance $t+1$. The output $o_1^e(t)$ is further processed by a stack of linear ReLU-activated layers to generate the final position estimate $\hat{\mathbf{p}}_t$. For static localization, the estimator is only evaluated at the last step $t=T$ of the horizon. }
    \label{fig:estimator_viz}
\end{figure}

Let $\bar{\boldsymbol{y}}(t)\triangleq[\Re\{y(t)\}, \Im\{y(t)\}]$. The \textit{policy} NN takes its hidden state vector $\boldsymbol{\ell}_{\mathcal{P}}(t)$ and the most recent observation $\bar{\boldsymbol{y}}(t)$ as inputs\footnote{Alternatively, the received signal strengths (RSSs) (i.e., $|y(t)|^2$ $\forall t$~\cite{active_sensing_ref1,active_sensing_ref2}) can be used as inputs to the policy NN at the BS. It will be shown later on, in the results' section, that our method performs well also with this input.} to its LSTM, producing the output $\boldsymbol{o}^p_1(t+1)$. This output then passes through an additional NN of linear ReLU-activated layers, yielding the vector $\boldsymbol{o}^p_2(t+1) \in \mathbb{R}^{N_{\rm ris}\rm{card}(\Theta) +1}$. Its first $N_{\rm ris}\rm{card}(\Theta)$ elements are passed through an element-wise ${\rm softmax}(\cdot)$ function to define a probability distribution over the configuration of the RIS elements, while the last element is transformed using a ${\rm tanh}(\cdot)$ and a ${\rm sign}(\cdot)$ functions to produce the single-bit $b(t)$-value to be transmitted\footnote{The proposed \textit{policy} NN can be easily modified to account for multi-bit $b(t)$-value messages: the network's output needs to be increased according to the total number of feedback bits, and its activations need to be replaced by a ${\rm softmax}(\cdot)$ function.} to the UE. On the other hand, the \textit{estimator} NN, which is depicted in Fig.~\ref{fig:estimator_viz}, 
processes $\bar{\mathbf{y}}(t)$ with its LSTM unit using the most recent hidden state $\boldsymbol{\ell}_{\mathcal{E}}(t)$, producing the output $\boldsymbol{o}^e_1(t)$. This output is then passed through a stack of linear layers activated by Rectified Linear Unit (ReLUs), resulting in an output $\hat{\boldsymbol{p}}_t \in \mathbb{R}^3$ that represents the estimation for the UE position at time instance $t$. In the case of localization, where $\mathbf{p}_t=\mathbf{p}$ is fixed, $\mathcal{E}$ is employed only once at the end of the episode. The observations are sequentially passed through the same LSTM weights, and the final (i.e., at $t=T$) output $\mathbf{o}_1^e(T)$ is provided  to the output stack of linear ReLU-activated layers to infer $\hat{\mathbf{p}}$.
We use the following notation to highlight the parameterization of the functions $g_{\rm bs}(\cdot)$ and $f(\cdot)$ appearing in expressions~\eqref{eq:twoPolicyDef_a} and \eqref{eq:decDEf}, respectively, from the respective RNNs $\mathcal{P}$ and $\mathcal{E}_I$:
\begin{align*}
    &g_{\rm bs}\left(\bar{\mathbf{y}}(t),\boldsymbol{\ell}_{\mathcal{P}}(t); \mathcal{P} \right), 
    \,\,f \left(\bar{\mathbf{y}}(t),\boldsymbol{\ell}_{\mathcal{E}_I}(t);\mathcal{E}_I \right).
\end{align*}
Notice that, instead of requiring as input the entire observation sequence (as in expressions~(\ref{eq:twoPolicyDef}) and (\ref{eq:decDEf})), only the most recent observation along with the hidden vector are needed, with the latter compressing all important past information. For long tracking sequences, processing the entire observation history at each instance $t$ can become computationally infeasible. 
\subsubsection{The UE Agent} An additional LSTM, referred to as the \textit{power} NN, $\mathcal{M}$, is maintained at the UE side. This NN processes the most recent control value $b(t)$ (received through the feedback channel) along with its hidden state $\boldsymbol{\ell}_{\mathcal{M}}(t)$ to select the power level $P(t+1)$ for the next time instance $t+1$. Thanks to the representational power of LSTMs, this NN can learn highly effective power control policies despite the strict constraint of the $1$-bit feedback message. As we will demonstrate later on, the performance of this decentralized scheme can approach that of systems with much richer feedback. 
We, henceforth, represent the parameterization of the function $g_{\rm ue}(\cdot)$ appearing in~\eqref{eq:twoPolicyDef_b} from the NN $\mathcal{M}$ as:
\begin{align*}
g_{\rm ue}(b(t), \boldsymbol{\ell}_{\mathcal{M}}(t); \mathcal{M}).
\end{align*}

\subsection{Proposed Training Procedure}
The training of the proposed DA approach comprises the following three stages. First, an initial estimator NN, $\mathcal{E}_I$, is trained on a dataset of randomly generated episodes (i.e., with random policies). Then, the policy, $\mathcal{P}$, and power, $\mathcal{M}$, NNs are evolved to collect UE trajectories that aid $\mathcal{E}_I$'s inference capabilities. In the sequel, when the policies have been learned, a final estimator NN $\mathcal{E}$ is retrained using data collected under the learned policies of the BS ($\mathcal{P}$) and UE ($\mathcal{M}$) agents. The first and last steps involve classic supervised learning on the MSE loss function. The key step is the intermediate stage, where the NNs $\mathcal{P}$ and $\mathcal{M}$ need to be jointly optimized. 
\subsubsection{Fitness Function}
We proceed by treating the pair $(\mathcal{P}, \mathcal{M})$ as a single optimization variable within the NE framework~\cite{eaht}. More specifically, each individual in the population is represented by a concatenated parameter vector:
\begin{equation}
\boldsymbol{i} \triangleq [\boldsymbol{w}_{\mathcal{P}}, \boldsymbol{w}_{\mathcal{M}}],
\end{equation}
and the fitness function is defined as follows: 
 \begin{equation}
        \label{eq:fitness}
    q(\boldsymbol{i}) =
    \begin{cases}
        -\hat{\mathbb{E}}_{\mathcal{P,M}}\left[\overset{T}{ \underset{t=1}{\sum}}
 P(t)\right], \quad  \text{if } \hat{\mathbb{E}}_{\mathcal{P,M}}\left[\overset{T}{ \underset{t=1}{\sum}}
 P(t)\right] > B_P \\
        - \frac{1}{T} \hat{\mathbb{E}}_{\mathcal{P,M},\mathcal{E}_I} \left[ \sum_{t=1}^T \|\hat{\mathbf{p}}_t - \mathbf{p}_t\|_2 \right], \quad \text{otherwise.}     \end{cases}\!\!\!\!\!\!\!,
        \end{equation}
where $\hat{\mathbb{E}}_X[\cdot]$ represents sample averaging, with respect to the random variable or set of variables $X$, over a large number $N_{\rm EP}$ of Monte Carlo episodes. It is noted that, in our application, besides the noise $n(t)$ in~\eqref{eq:received_signal}, the stochastic policies and the estimator affect the averaging. Intuitively, the fitness function in~\eqref{eq:fitness} penalizes individuals that fail to satisfy the power budget constraint $B_P$. Among those that satisfy it, individuals leading to observations of higher quality with smaller tracking errors are preferred. Note also that, in static localization objectives, the second branch of the fitness function in~(\ref{eq:fitness}) 
simplifies to $\hat{\mathbb{E}}_{\mathcal{P},\mathcal{M},\mathcal{E}_I}
\left[\|\hat{\mathbf{p}}_t - \mathbf{p}_t\|_2\right]
$.

For each evaluation episode, a random UE position $\mathbf{p}_0$ is sampled along with related parameters concatenated in the state $\mathbf{x}_0$, and the hidden states of all NNs are initialized. At each $t$-th time instance, the following four operations occur: 
 \begin{itemize}
 \item The UE state $\mathbf{x}(t)$ is updated according to the motion model $m(\cdot)$, as described in the state model in~(\ref{eq:target_state_update}).
     \item An observation $y(t)$ is sampled from~\eqref{eq:received_signal} using $P(t)$ and $\boldsymbol{\Phi}(t)$ which is provided to the policy NN $\mathcal{P}$.
     \item  The latter network outputs $\boldsymbol{\Phi}(t+1)$ and an $1$-bit message $b(t)$. On the UE side, the power NN $\mathcal{M}$ uses $b(t)$ along with its hidden state to select $P(t+1)$.
     \item The vector  $\bar{\boldsymbol{y}}(t)$ is also provided to the  initial estimator NN $\mathcal{E}_I$ to compute the estimation error, which determines the individual’s fitness.
 \end{itemize}
 
\subsubsection{Optimization Procedure}
The proposed optimization process is performed by first initializing a population of $L_{\rm pop}$ random individuals $\boldsymbol{i}_1,\ldots,\boldsymbol{i}_{L_{\rm pop}}$, with $\boldsymbol{i}_\ell \triangleq [\boldsymbol{w}_{\mathcal{P},\ell}, \boldsymbol{w}_{\mathcal{M},\ell}]$ $\forall \ell=1,\ldots,L_{\rm pop}$.  These individuals are updated over multiple optimization iterations referred to as generations.
 For each generation within the range $1,\ldots,N_{\rm gen}$, the evolutionary process iterates through the following steps (see also Algorithm~\ref{alg:ne_training} for more details): \begin{itemize}
     \item  \textbf{Evaluation:} For every individual $\boldsymbol{i}_\ell$ in the current population, the corresponding parameter vector is first split to reconstruct the weight matrices of the BS policy NN and the UE power NN.  The individual's fitness is calculated and stored in order to select the best performing individuals for the population update.
     \item \textbf{Selection:} Once all individuals have been evaluated, the population is sorted in descending order based on their computed fitness scores. To promote the propagation of high quality traits, only the top performing fraction of the population, specifically the best $\lfloor{L_{\rm pop}/4}\rfloor$ individuals, are selected to survive and, consequently, serve as parents for the next generation.
     \item  \textbf{Crossover:} To replenish the population back to its full size $L_{\rm pop}$, offspring are generated by mating the selected parents. During this operation, the weight vectors of two randomly chosen parents are merged to construct a new candidate solution. Specifically, we employ a uniform crossover strategy: the offspring's weight vector is initialized to zero, and each individual parameter is then assigned the value of the corresponding weight from one of the two parents, selected with equal probability (i.e., a 50\% chance for each parent).
     \item \textbf{Mutation:} To prevent premature convergence to local optima, a mutation operator is applied to the newly generated offspring. This involves adding zero-mean Gaussian noise with standard deviation $\sigma_{\rm mut}$ to each component $c$ of $\ell$'s weight vector with a probability $p_{\rm mut}$. Formally, for the $\ell$-th individual, the mutation operator  applied to each component $c$ is:
     \begin{equation}\label{eq:mutation_def}
     \boldsymbol{i}_{\ell}[c] \gets \begin{cases}   \boldsymbol{i}_{\ell}[c] + \mathcal{N}(0,\sigma_{\rm mut}^2) \quad \text{with probability } p_{\rm mut} \\
     \boldsymbol{i}_\ell[c]  \quad \text{with probability } 1- p_{\rm mut}
      \end{cases}.
     \end{equation}
 \end{itemize}
\begin{algorithm}[!t]
\caption{Proposed DA Active Sensing Training}
\label{alg:ne_training}
\begin{algorithmic}[1]
\renewcommand{\algorithmicrequire}{\textbf{Input:}}
\renewcommand{\algorithmicensure}{\textbf{Output:}}
\REQUIRE Population size $L_{\rm pop}$, number of generations $N_{\rm gen}$, evaluation episodes $N_{\rm EP}$, mutation probability $p_{\rm mut}$ and standard deviation $\sigma_{\rm mut}$, and horizon $T$.
\ENSURE Optimized policy and estimator NNs at BS as well as power NN at UE.

\STATE \textit{Stage 1: Initial Estimator Training}
\STATE Create dataset $\mathcal{D}_{\rm rand}$ using random RIS phase profiles $\Phi(t)$ and power levels $P(t)$ at each time instant $t$.
\STATE Train initial estimator NN $\mathcal{E}_I$ on $\mathcal{D}_{\rm rand}$ for minimizing the MSE in $\mathcal{OP}$.

\STATE \textit{Stage 2: Optimize policies}
\STATE Initialize population $ \{\boldsymbol{i}_\ell = [\boldsymbol{w}_{\mathcal{P},\ell}, \boldsymbol{w}_{\mathcal{M},\ell}]\}_{\ell=1}^{L_{\rm pop}}$.
\FOR{generation $ 1$ to $N_{\rm gen}$}
    \FOR{individual $\ell = 1$ to $L_{\rm pop}$}
        \STATE Convert individual $\boldsymbol{i}_\ell$ into NNs $\mathcal{P}_\ell$ and $\mathcal{M}_\ell$.
        \FOR{episode $e = 1$ to $N_{\rm EP}$}
            \STATE Sample user location $\mathbf{p}_0 \in \mathcal{P}$ and channel realizations $h_d^e,\mathbf{h}^e_{\rm bs,ris},\mathbf{h}^e_{\rm ris,ue}$.
            \STATE Initialize LSTM hidden state vectors: $\boldsymbol{\ell}_{\mathcal{P}_\ell}(0)$, $\boldsymbol{\ell}_{\mathcal{M}_\ell}(0)$, and $\boldsymbol{\ell}_{\mathcal{E}_I}(0)$. 
            \FOR{$t = 1$ to $T$}
            \STATE Move UE according to eq. (\ref{eq:target_state_update}) and compute the new channels.
            \STATE Compute $y(t)$ according to eq. (\ref{eq:received_signal}).
                \STATE \textit{BS policy NN $\mathcal{P}_\ell$:} $[\Phi(t+1), b(t)] \leftarrow g_{\rm bs}
                \left(\bar{\mathbf{y}}(t),\boldsymbol{\ell}_{\mathcal{P}_\ell}(t); \mathcal{P}_\ell \right)$.
                \STATE \textit{UE power NN $\mathcal{M}_\ell$:} $P(t+1) \leftarrow g_{\rm ue}(b(t), \boldsymbol{\ell}_{\mathcal{M}_\ell}(t); \mathcal{M}_\ell)$.
                \STATE \textit{BS Estimation NN $\mathcal{E}_I$:}
                $\mathbf{\hat{p}}_t =f\left(\bar{\mathbf{y}}(t),\boldsymbol{\ell}_{\mathcal{E}_I}(t);\mathcal{E}_I \right)
                $
                \STATE Store the error $\|\hat{\mathbf{p}}_t-\mathbf{p}_t \|_2$.
            \ENDFOR
            \STATE Calculate the fitness of the $e$-th episode, $q^e(\boldsymbol{i}_\ell)$, via~(\ref{eq:fitness}).
        \ENDFOR
        \STATE Compute the final fitness:
        \hspace{0.5cm} $q(\boldsymbol{i}_\ell) \leftarrow \frac{1}{N_{\rm EP}} \sum_{e=1}^{N_{\rm EP}} q^e(\boldsymbol{i}_\ell)$.
    \ENDFOR
    \STATE Sort the population based on fitness values $\{q(\mathbf{i}_\ell)\}_{\ell=1}^{L_{\rm pop}}$.
    \STATE Select elite parents $\mathcal{P}_{\rm elite} \leftarrow \{\boldsymbol{i}_\ell\}_{\ell=1}^{\lfloor L_{\rm pop}/4 \rfloor}$.
    \STATE \underline{Crossover:} Refill the population by merging pairs from $\mathcal{P}_{\rm elite}$ (uniform selection).
    \STATE \underline{Mutate:} Perturb the new individuals via eq. (\ref{eq:mutation_def}).
\ENDFOR
\STATE Extract $\mathcal{P}_\ell^*$ and $\mathcal{M}_\ell^*$ corresponding to the individual  $\boldsymbol{i}_\ell^*$ with the highest fitness from the resulting (final) population.

\STATE \textit{Stage 3: Fine Tuning}
\STATE Collect dataset $\mathcal{D}_{\rm opt}$ using evolved policies $\mathcal{M}_\ell^*$ and $\mathcal{P}_\ell^*$.
\STATE \textit{BS Estimation NN $\mathcal{E}$:} Retrain final estimator NN on $\mathcal{D}_{\rm opt}$ to minimize the MSE in $\mathcal{OP}$.
\RETURN $\mathcal{P}_\ell^*$, $\mathcal{M}_\ell^*$, and $\mathcal{E}$.
\end{algorithmic}
\end{algorithm}
\subsubsection{Computational Complexity}
Let $T_{\rm FP}$ denote the computational time required for a forward pass of the policy and power NNs at the BS and UE, respectively. Given their structural similarity, we assume identical inference latency for both of these two agents. Consequently, evaluating the fitness function of the entire population entails a computational complexity of $\mathcal{O}(L_{\rm pop} N_{\rm EP} T T_{\rm FP})$. The associated genetic operations (i.e., crossover and mutation) scale linearly with the parameter space, contributing complexity of $\mathcal{O}(L_{\rm pop} W)$, where $W$ represents the total number of learnable weights (i.e., the total number of  parameters of both the policy ($\mathcal{P}$) and power ($\mathcal{M}$) NNs). Additionally, ranking the population based on fitness requires $\mathcal{O}(L_{\rm pop} \log(L_{\rm pop}))$ of complexity. Aggregating these components over $N_{\rm gen}$ generations yields the final total complexity: $$\mathcal{O}\left( N_{\rm gen} \left( L_{\rm pop} N_{\rm EP} T T_{\rm FP} + L_{\rm pop} W + L_{\rm pop} \log(L_{\rm pop})\right) \right).$$
\subsection{Extension to Multi-Antenna Base Stations}\label{sec:multi_antenna_extension}
\begin{figure}
    \centering
    \includegraphics[width=\linewidth]{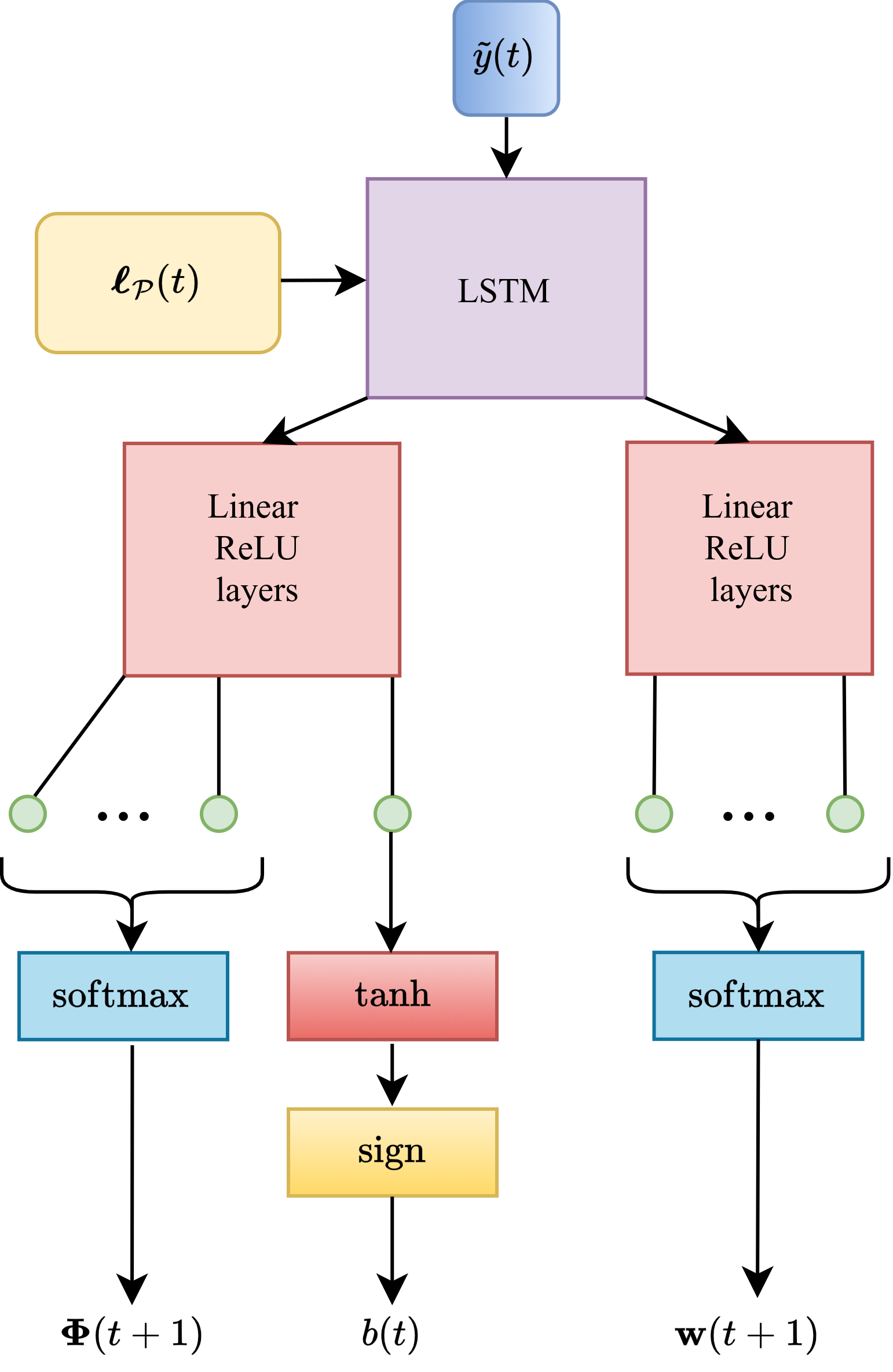}
    \caption{The proposed policy NN for the multi-antenna BS case. An LSTM core component extracts the important temporal features from the observation sequence. The LSTM's output is then processed by two separate FF stacks: the one stack is in charge of selecting the RIS profile and the binary message (similar to the single-antenna case), while the other second stack is configured to output an $N_{\rm bs}$-dimensional vector. Application of the ${\rm softmax}(\cdot)$ function to this output yields a valid beamforming selection, i.e., such that $\mathbf{w}(t) \in \mathcal{W}$.}
    \label{fig:MIMO_viz}
\end{figure}
The proposed DA active sensing framework for UE localization and tracking can be easily extended to a multi-antenna BS, which will now also be capable of dynamically adjusting its beamforming operation for further boosting the estimation accuracy. To this end, the following modifications on the policy NN $\mathcal{P}$ at an $N_{\rm bs}$-antenna BS are needed. Let $\mathbf{y}(t) \in \mathbb{C}^{N_{\rm bs} \times 1}$ denote the vector form of the baseband received signal at the BS side, which can be expressed as:
\begin{equation}
    \mathbf{y}(t)\triangleq\sqrt{P(t)} \left( \mathbf{h}_d+\mathbf{H}_{\rm bs,ris} \boldsymbol{\Phi}(t) \mathbf{h}_{\rm ris, ue}\right) x(t)+\mathbf{n}(t),
\end{equation}
where $\mathbf{h}_d\in\mathbb{C}^{N_{\rm bs}\times 1}$ and $\mathbf{H}_{\rm bs,ris}\in\mathbb{C}^{N_{\rm bs}\times N_{\rm ris}}$ represent the gains of the direct BS-UE and the BS-RIS channels, respectively. Note that, similar to~(\ref{eq:received_signal}), all involved channel vectors and matrices depend on the unknown UE coordinates $\mathbf{p}_t$. In addition, $\mathbf{n}(t)\sim\mathcal{N}(\mathbf{0}_{N_{\rm bs}},\sigma^2_n\mathbf{I}_{N_{\rm bs}})$ denotes the AWGN. Consider that, at each time instance $t$, a digital combining vector $\mathbf{w}(t) \in \mathbb{C}^{N_{\rm bs}\times1}$ is applied at BS's baseband as follows:
\begin{equation}
\tilde{y}(t)\triangleq\mathbf{w}^{\rm H}(t)\mathbf{y}(t),
\end{equation}
providing a complex-valued scalar. This post-processing scalar can be then plugged into the previously detailed single-antenna active sensing machinery. In practice, $\mathbf{w}(t)$ needs to be selected from a finite beam codebook $\mathcal{W}$. For example, a simplified version of the 3GPP 5GNR Type I codebook \cite{3GPP_TS_38.214_codebooks} considers $\mathcal{W}$ containing the rows of the $N_{\rm bs} \times N_{\rm bs}$ Discrete Fourier Transform (DFT) matrix. 

Following our active sensing formulation, the mapping $g_{\rm bs}(\cdot)$ of the multi-antenna BS will be now responsible for configuring the BS digital combiner and the RIS phase profiles, as well as for selecting the $1$-bit feedback messages, i.e.:
\begin{equation}\label{eq:twoPolicyDef_multiantenna}
  \{\boldsymbol{\Phi}(t+1),\mathbf{w}(t+1),b(t)\} =  g_{\rm bs}(\tilde{y}(1),\tilde{y}(2),\ldots,\tilde{y}(t)).
\end{equation}
To this end, the policy NN $\mathcal{P}$ is extended by adding one more output branch, as shown in Fig.~\ref{fig:MIMO_viz}. This branch receives as input the most recent LSTM output $\mathbf{o}_1^p(t+1)$ and then passes it through a linear ReLU-activated stack, whose final layer is configured to output an $N_{\rm bs}$ dimensional vector. In the sequel, a ${\rm softmax}(\cdot)$ is applied to this output in order to sample a combiner $\mathbf{w}(t+1) \in \mathcal{W}$ for the $(t+1)$-th time instance. All other aspects of the operation of the NNs and the procedure for solving $\mathcal{OP}$ remain unchanged.

\section{Numerical Results and Discussion}\label{sec:examples}
In this section, we evaluate the performance and robustness of the proposed DA active sensing framework through extensive simulations. Specifically, the effectiveness of the presented tracking scheme is investigated across various operational scenarios, encompassing both mobile UEs under diverse dynamic motion models, as well as static localization environments. To rigorously assess the merits of our algorithm, its performance has been benchmarked against both classical estimation techniques~\cite{EKF-ref1,PF1,fingerprint} as well as learning-based sensing agents~\cite{activeSensingLocJournal}. Before detailing the simulation results, we first present the core algorithmic hyperparameters employed throughout all our following experiments.

The parameters of the CoSyNE algorithm~\cite{cosyne} are set to 
$p_{\rm mut}=\sigma_{\rm mut}=0.5$, $L_{\rm pop}=50$, and $N_{\rm gen}=100$. The policy NN's LSTM was chosen to have $2$ hidden layers of $512$ units, and each of the FF branches for the selection of the RIS phase profile and the $1$-bit power parameter message was chosen to have a single hidden layer of $128$ and $32$ units, respectively. The power NN’s LSTM had also $2$ hidden layers of $512$ units, followed by a linear layer of $64$ hidden units. The estimator NN was designed to have a similar LSTM, followed by $2$ ReLU-activated hidden layers each of $128$ units. Both initial and final estimators were trained on $50000$ sequences. 

\begin{figure*}[t!]
    \centering
    \begin{subfigure}[b]{0.32\textwidth}
        \centering
        \includegraphics[width=\textwidth]{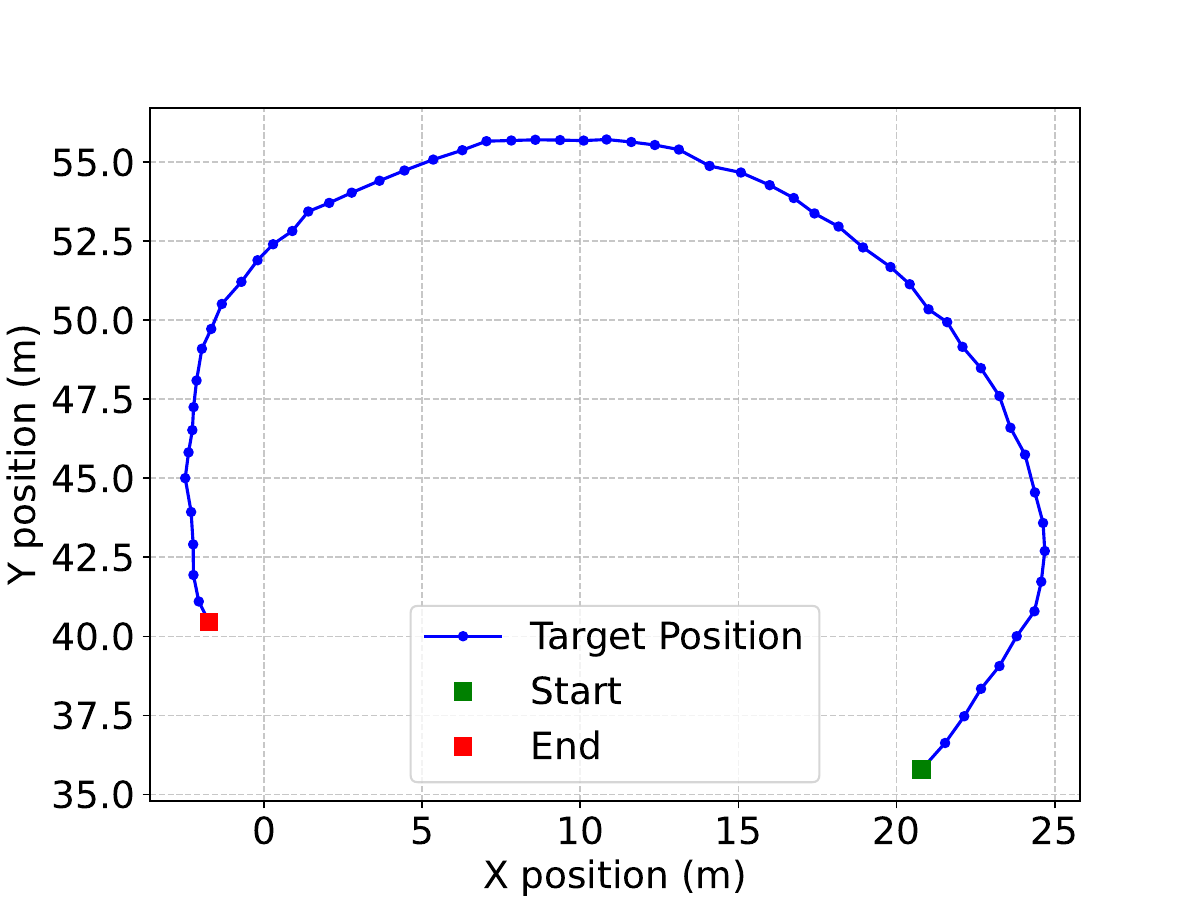}
        \caption{$\omega_0=0.2$}
        \label{fig:smallomenga}
    \end{subfigure}
    \hfill
    \begin{subfigure}[b]{0.32\textwidth}
        \centering
        \includegraphics[width=\textwidth]{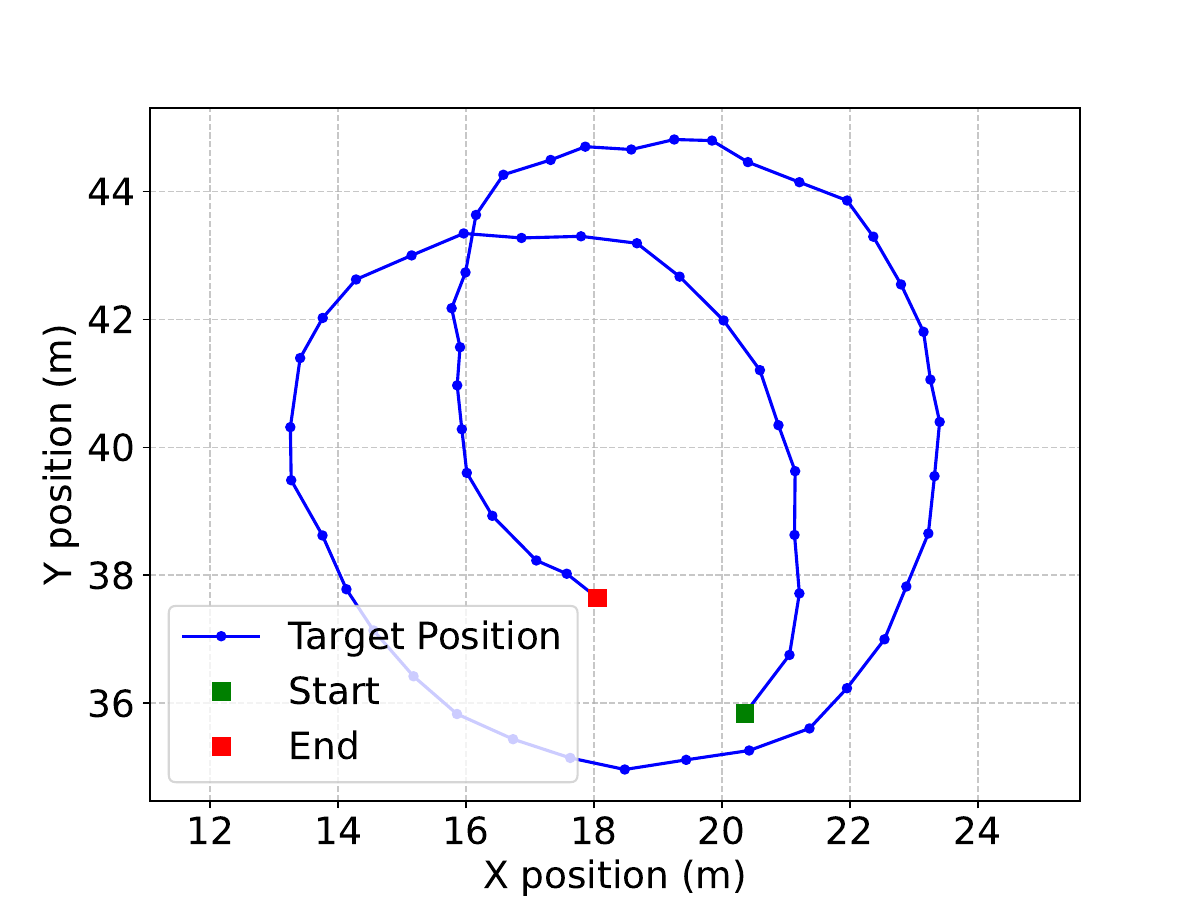}
        \caption{$\omega_0=0.4$}
        \label{fig:midomega}
    \end{subfigure}
    \hfill
    \begin{subfigure}[b]{0.32\textwidth}
        \centering
        \includegraphics[width=\textwidth]{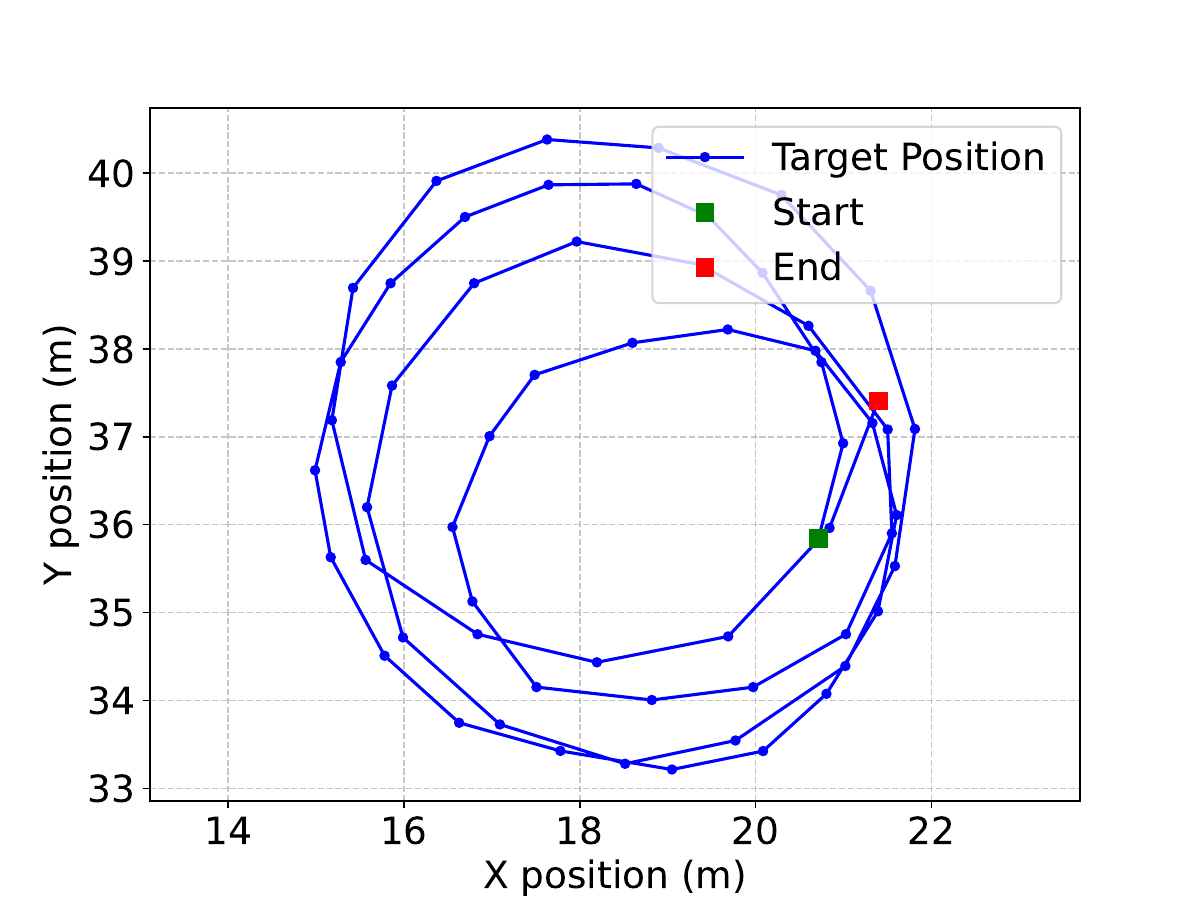}
        \caption{$\omega_0=0.8$}
        \label{fig:largeomega}
    \end{subfigure}
    \caption{The considered example UE trajectories for different initial turn rates $\omega_0$, considering the motion variance $\sigma_{\rm ue}=0.1$. It can be observed that a low initial turn rate (i.e., $\omega_0=0.2$ in (a)) yields a relatively smooth trajectory, whereas increasing this rate to $0.4$ (as in (b)) results in more aggressive maneuvers. Finally, the turn rate of $\omega_0=0.8$ in (c) induces highly rapid directional changes.}
    \label{fig:tracking_traj_examples}
\end{figure*}
\subsection{System Setup}
A carrier frequency of $5$~GHz has been considered, corresponding to a signal wavelength of $\lambda=6$~cm. The BS was assumed located at the position $[40,-40,10]$~m, whereas the top-left unit element of the RIS was placed at the origin $[0,0,0]$~m. The metasurface was modeled as a uniform rectangular array, placed parallel to the $x$-$z$ plane, consisting of $N_{\rm ris}=400$ elements with binary phase responses, unless otherwise indicated. Those elements were assumed the standard half-wavelength element spacing of $ \lambda/2 = 3$~cm. To model a typical indoor propagation environment, we adopted a log-distance path loss model with an exponent of $ 2.5$; this specific value is empirically justified for sub-6 GHz indoor systems experiencing NLoS obstructions~\cite{IndoorPathLossExp}. Alongside the path loss, we considered Ricean fading conditions with a $\kappa=10$ dB factor, unless otherwise indicated, and applied an extra attenuation of $10$~dB to the direct BS-UE channel. Furthermore, assuming a transmission bandwidth of $10$~MHz and a background noise power spectral density of $-170$~dBm/Hz, the effective noise level at the BS side $\sigma_{{n}}^2$ was set to $-100$~dBm~\cite{activeSensingLocJournal}. Finally, the maximum UE pilot transmission power in the uplink was bounded at $P_{\rm max}=30$~dBm, and the $T$-horizon power constraint in the $\mathcal{OP}$ formulation was set as $B_{P}=0.5\sum_{t=1}^TP_{\rm max}$.

A two-dimensional motion model was considered, where the UE moves along the $x$ and $y$ axes under a nonlinear Coordinated Turn (CT) model. Physically, this model captures the realistic maneuvering behavior of a mobile user, such as navigating a curve or altering its path. It assumes that during any short sampling interval, the user maintains a constant forward speed while turning at a steady rate. The entire state was a $5$-dimensional vector $\mathbf{x}(t)=[p_{x,t}, p_{y,t}, v_{x,t}, v_{y,t}, \omega_t]^T$, where $p_{x,t}$ and $p_{y,t}$ represent the Cartesian position, $\mathbf{v}_t\triangleq[v_{x,t},v_{y,t}]$ denotes the velocity, and $\omega_t$ is the instantaneous turn rate. 

Depending on the turn rate, the state transition function $m(\mathbf{x}(t))$ in~\eqref{eq:target_state_update} operates in two regimes. If $\omega_t \neq 0$, the UE moves along a circular arc with a constant angular velocity over the sampling interval, and $m(\mathbf{x}(t))$ is equal to:
$$\begin{bmatrix}
p_x(t) + \frac{v_x(t)}{\omega(t)} \sin(\omega(t) \Delta T) - \frac{v_y(t)}{\omega(t)} (1 - \cos(\omega(t) \Delta T)) \\
p_y(t) + \frac{v_x(t)}{\omega(t)} (1 - \cos(\omega(t) \Delta T)) + \frac{v_y(t)}{\omega(t)} \sin(\omega(t) \Delta T) \\
v_x(t) \cos(\omega(t) \Delta T) - v_y(t) \sin(\omega(t) \Delta T) \\
v_x(t) \sin(\omega(t) \Delta T) + v_y(t) \cos(\omega(t) \Delta T) \\
\omega(t)
\end{bmatrix},$$ 
where the sampling interval was set to $\Delta T=0.5$ sec, and the initial turn rate $\omega_0$ was sampled uniformly from $[-\omega_{\rm max}, \omega_{\rm max}]$. As visualized in Fig.~\ref{fig:tracking_traj_examples}, the initial turn rate $\omega_0$ significantly dictates the curvature and aggressiveness of the resulting trajectory/episode. However, this standard CT formulation introduces a mathematical singularity (division by zero) when the target stops turning and moves in a straight line. To ensure numerical stability and model continuity, we have evaluated the limit as $\omega_t \to 0$; this seamlessly reduced the system to the following state update model:
$$\lim_{\omega_t \to 0} m(\mathbf{x}(t)) = \begin{bmatrix}
p_x(t) + v_x(t) \Delta T \\
p_y(t) + v_y(t) \Delta T \\
v_x(t) \\
v_y(t) \\
\omega(t)
\end{bmatrix}.$$

\subsection{Results for Tracking}
For the UE tracking investigations, we have also simulated a supervised tracking LSTM and two popular filters (all with random policies): the Extended Kalman Filter (EKF)~\cite{EKF-ref1,EKF-ref2} and the Particle Filter (PF)~\cite{PF1,PF2} with $500$ particles. 
In addition, acknowledging DRL's adoption for other active sensing problems (e.g., ~\cite{active_sensing_ref2}) as well as for RIS phase configuration control~\cite{AlexandroPervasive}, we designed a benchmark based on the popular and powerful Advantage Actor Critic (A2C) algorithm~\cite{A2C}. To this end, we followed a similar three-step hybrid training approach to our proposed DA active tracking framework, by only replacing the implementation of the second step with a single A2C agent being in charge of selecting both the RIS phase profile and the UE power level. To enforce the power constraint for the latter, a Lagrangian reward was used similar to~\cite{SafeDRL}. For fairness, this agent utilized recurrent actor and critic NNs with similar structure and size as our NE-optimized policies. Training consisted of $50000$ episodes using the default A2C hyperparameters of a popular open source implementation provided in~\cite{stable-baselines3}. It is noted that, by comparing against this method, the effectiveness of the proposed NE procedure over off-the-shelf DRL algorithms can be quantified. Overall, we conducted experimental investigations with respect to the following system parameters: 
\begin{itemize}
    \item \textbf{Effect of motion noise:} The maximum initial turn rate was set to $\omega_{\max}=0.2$, the horizon  to $T=20$, and the motion variance $\sigma_{\rm ue}$ was varied from $0.1$ to $0.4$.
    \item \textbf{Effect of angle rate initialization:} Parameter $\omega_{\max}$ was varied from $0.2$ to $0.8$ in order to verify the ability of our model to generalize to diverse turning motions. The horizon was fixed to $T=20$ and the noise to $\sigma_{\rm ue}=0.4$.
    \item \textbf{Effect of the tracking horizon:} 
    The noise was fixed to $\sigma_{\rm ue}=0.4$, $\omega_{\max}$ was set to $0.2$, and the time horizon was varied from $T=40$ to $70$.
\end{itemize}

\begin{figure*}[t!]
    \centering
    \begin{subfigure}[b]{0.32\textwidth}
        \centering
        \includegraphics[width=\textwidth]{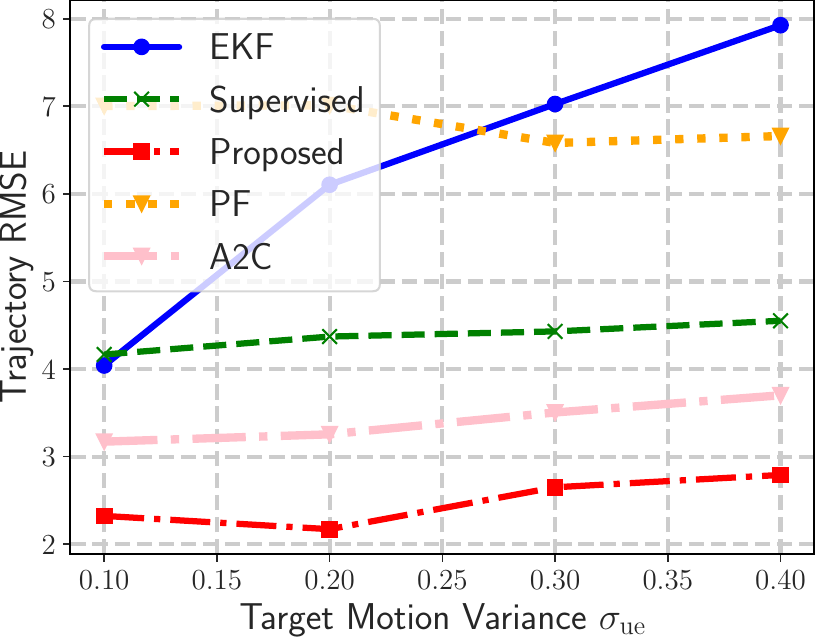}
        \caption{The effect of motion variance $\sigma_{\rm ue}$ for $\omega_{\max}=0.2$ and $T=20$.}
        \label{fig:tracknoise}
    \end{subfigure}
    \hfill
    \begin{subfigure}[b]{0.32\textwidth}
        \centering
        \includegraphics[width=\textwidth]{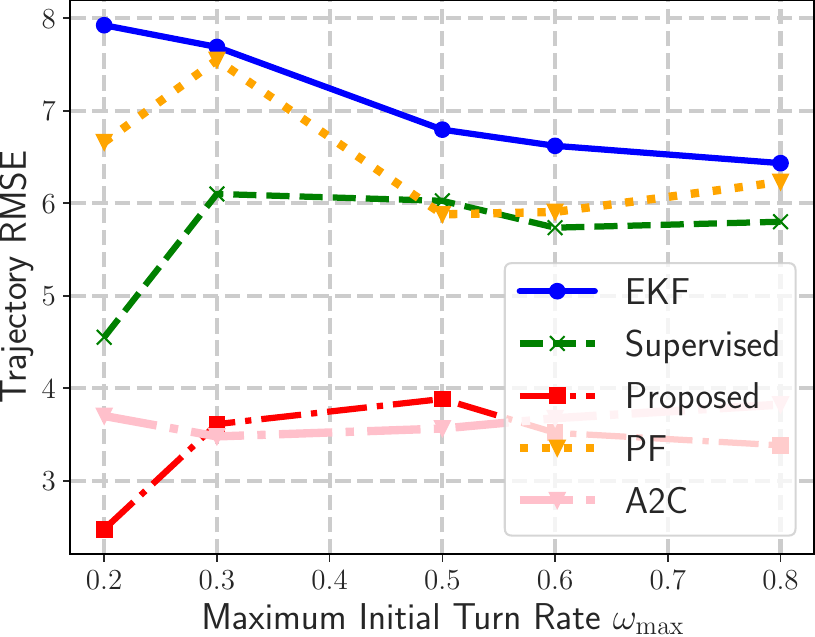}
        \caption{The effect of $\omega_{\max}$ for $\sigma_{\rm ue}=0.4$ and $T=20$. }
        \label{fig:trackangle}
    \end{subfigure}
    \hfill
    \begin{subfigure}[b]{0.32\textwidth}
        \centering
        \includegraphics[width=\textwidth]{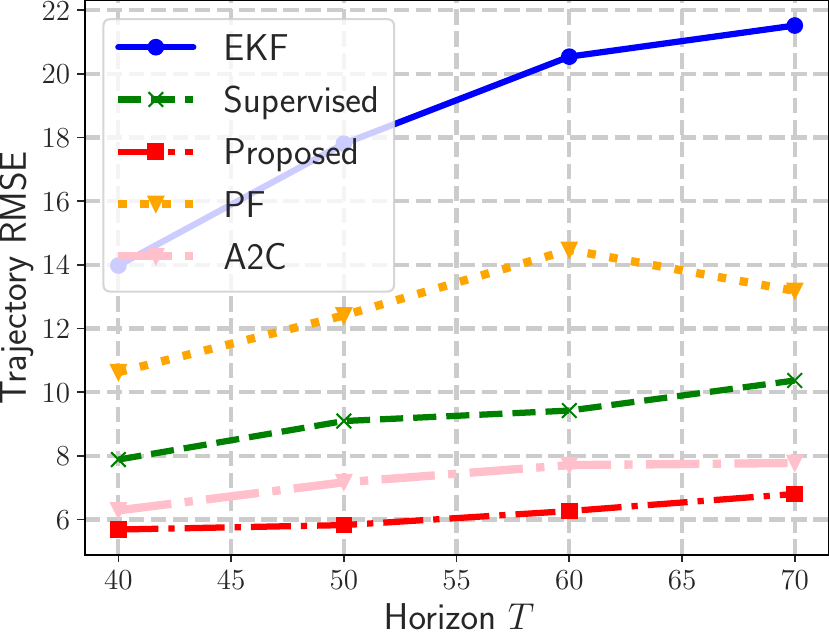}
        \caption{The effect of tracking horizon $T$ for $\sigma_{\rm ue}=0.4$ and $\omega_{\max}=0.2$.}
        \label{fig:trackhorizon}
    \end{subfigure}
    \caption{Trajectory RMSE performance over various UE motion and system parameters.}
    \label{fig:trackingResults}
\end{figure*}

\begin{figure}
    \centering
    \includegraphics[width=0.75\linewidth]{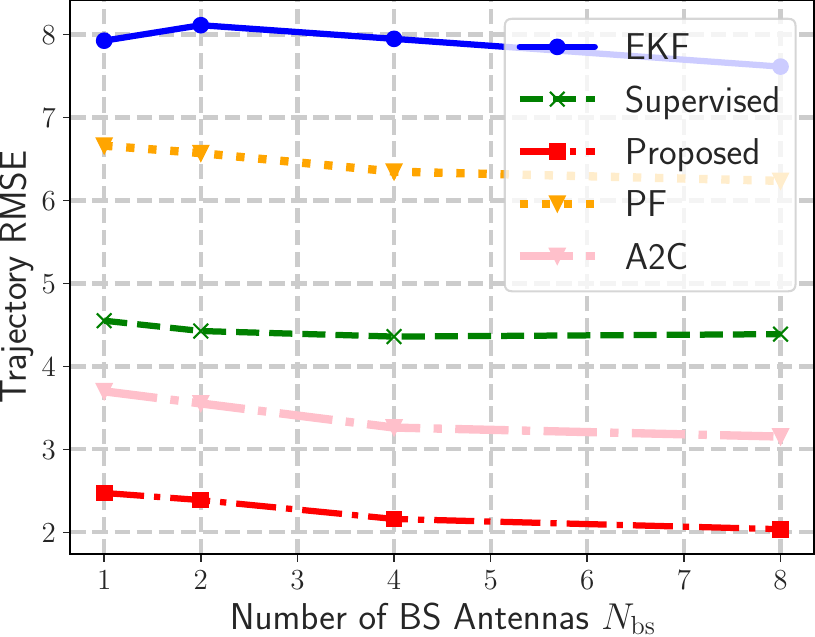}
    \caption{Trajectory RMSE performance versus the number of BS antennas $N_{\rm bs}$, considering $\omega_{\max}=0.2$, $\sigma_{\rm ue}=0.4$, and $T=20$.}
    \label{fig:ma_results}
\end{figure}

Figure~\ref{fig:trackingResults} illustrates the  Root Mean Squared Error (RMSE) across a range of UE motion dynamics and system parameters, with the provided results being averaged over $1000$ independent \textcolor{black}{episodes/trajectories} to ensure a statistically robust and fair evaluation of each algorithm's performance. As shown in the figure, the proposed DA active sensing scheme consistently outperforms all baseline methods. Notably, it demonstrates strong robustness against increasing motion noise $\sigma_{\rm ue}$, higher maximum turn rates $\omega_{\max}$, and longer tracking horizons $T$. It is also shown that the A2C benchmark outperforms traditional schemes with random policies, without, however, being able to match the performance of our proposed method, signifying the importance of our NE approach.
Furthermore, Fig.~\ref{fig:ma_results} investigates the scalability of our algorithm for an increasing number of BS antennas $N_{\rm bs}$, where, in this case, the BS was configured as a uniform linear array. It is again shown that our scheme is superior to all benchmarks, with increasing $N_{\rm bs}$  resulting in slightly boosting tracking performance.

\subsection{Results for Localization}
In the localization experiments, the UE's static position was chosen as uniformly distributed inside the cubic area $[20 \pm 15, 20 \pm 20m,-20]$ m, modeling spatial uncertainty in the $x$ and $y$ axes~\cite{activeSensingLocJournal}. 
The performance of our active sensing scheme was compared against the following localization algorithms: \begin{itemize}
    \item \textbf{Fingerprinting} \cite{fingerprint}: 
    The sequence of the $T$ RIS phase profiles was predetermined, non-adaptive and random, and the UE pilot transmit power levels at each time instance $t$ were sampled uniformly in $[0,P_{\max}]$. Each $1\,\mathrm{m^2}$ block in the candidate UE location area was assigned a fingerprint sequence $|y(1)|^2, \ldots, |y(T)|^2$, precomputed and stored in a database. During operation, a $5$-nearest neighbor fingerprint classifier was employed. We also considered another variant of this scheme where the RIS phase profile was optimized using a genetic algorithm. The default implementation of genetic algorithms from a popular open source library has been used~\cite{gad2023pygad}.
    \item \textbf{Supervised learning}: We considered  a supervised NN trained on $70000$ sequences/episodes with random RIS phase profiles and power levels sampled uniformly in the range $[0,P_{\rm max}]$ at each time instance $t$. We used an FF model with $4$ hidden layers each with $400$ units. This NN stacks the entire observation sequence into a large vector, which is then mapped into the final $3$-dimensional estimate.
    \item \textbf{End-to-end backpropagation with RIS of continuous phase responses} \cite{activeSensingLocconf,activeSensingLocJournal}: A scheme comprising an intelligent active sensing agent that selects continuous-valued RIS phase profiles, and the UE always transmitting with power $P_{\rm max}$ has been simulated. This agent was equipped with an LSTM that decides $\boldsymbol{\Phi}(t)$ using the current observation $y(t)$ and its hidden states. The hidden states were then passed through an FF branch for the final position estimation. The structure of the LSTM layers were similar to those used in the proposed DA scheme. Training was conducted using end-to-end backpropagation on the considered MSE loss \cite{activeSensingLocconf,activeSensingLocJournal}.
    \item \textbf{Deep reinforcement learning:}  The second step of the proposed DA active sensing scheme, which is tasked with the joint selection of $\boldsymbol{\Phi}(t)$ and $P(t)$,  was implemented an A2C agent, as in the tracking comparisons.
\end{itemize} 

\begin{figure*}[t!]
    \centering
    \begin{subfigure}[b]{0.32\textwidth}
        \centering
        \includegraphics[width=\textwidth]{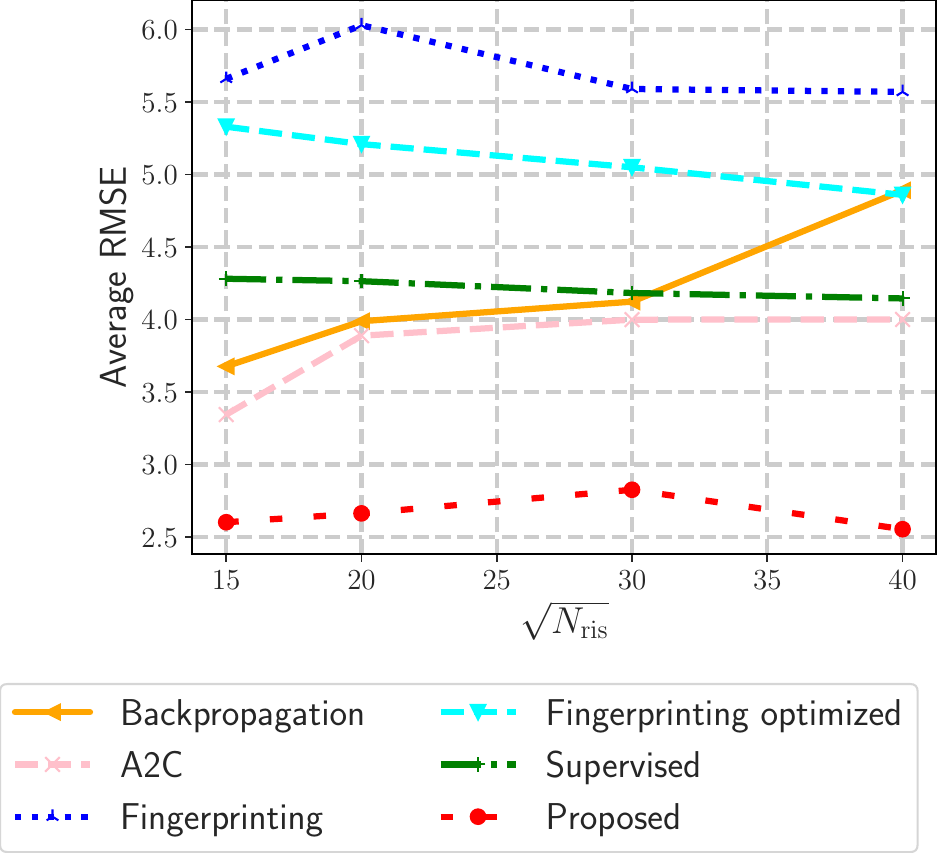}
        \caption{Effect of the RIS size $N_{\rm ris}$: the Ricean factor is set to $\kappa=10$ dB and the power of the noise level $\sigma_{ n}$ is set to $-100$ dBm~\cite{activeSensingLocJournal}.}
        \label{fig:ris_size}
    \end{subfigure}
    \hfill
    \begin{subfigure}[b]{0.32\textwidth}
        \centering
        \includegraphics[width=\textwidth]{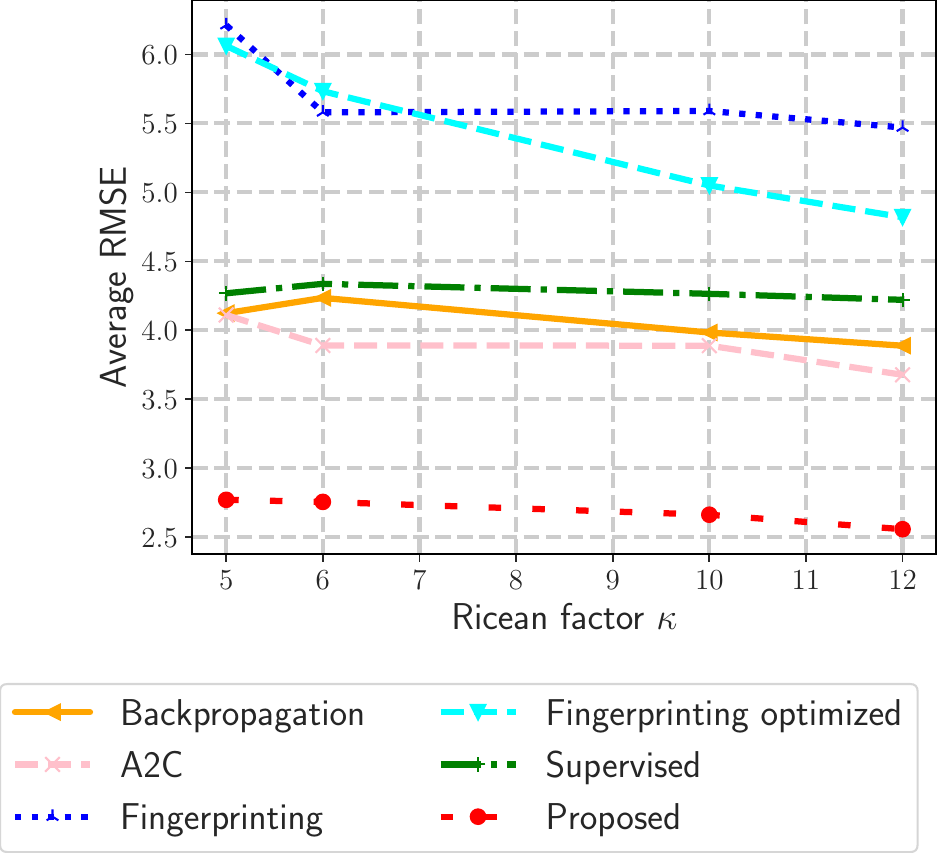}
        \caption{Effect of the Ricean factor $\kappa$: the RIS panel size is set to $N_{\rm ris}=400$, and the power of the  noise level $\sigma_{n}$ is set to $-100$ dBm.}
        \label{fig:ricean}
    \end{subfigure}
    \hfill
    \begin{subfigure}[b]{0.32\textwidth}
        \centering
        \includegraphics[width=\textwidth]{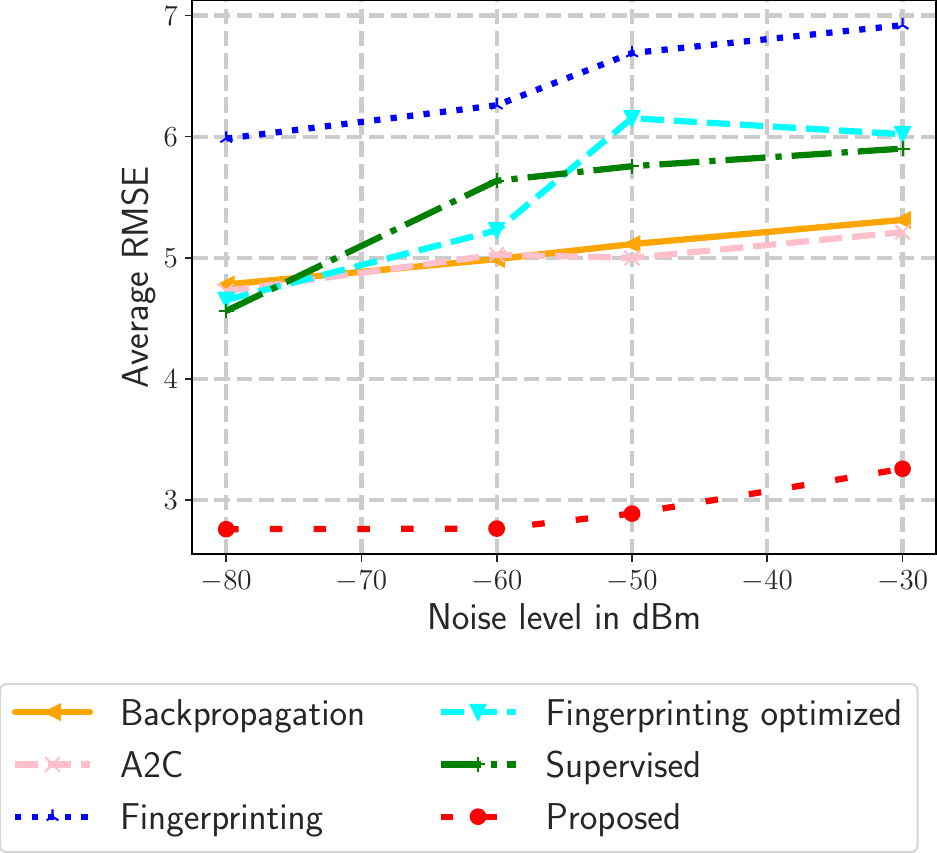}
        \caption{Effect of the observation noise $\sigma_{n}$: the RIS panel size is set to $N_{\rm ris}=400$ and the Ricean factor to $\kappa=10$ dB.}
        \label{fig:system_noise}
    \end{subfigure}
    \caption{Localization RMSE performance over various system parameters.}
    \label{fig:main_results}
\end{figure*}
Three sets of experiments, as demonstrated in Fig.~\ref{fig:main_results}, were conducted to evaluate the robustness and scalability of the proposed algorithm across a broad range of wireless configurations. The key findings are summarized as follows:

\begin{itemize}
    \item \textbf{Impact of the RIS size $(N_{\rm ris})$:} While larger RISs provide higher beamforming gain to steer signals toward the BS, they also expand the action space, increasing the number of decision variables and the complexity of the optimization (hence, requiring larger NNs). As depicted in Fig.~\ref{fig:ris_size}, our approach scales well with larger RISs. On the contrary, the gradient-based scheme of \cite{activeSensingLocJournal}  and A2C display decreasing performance with larger RISs.  
    
    \item \textbf{Effect of scattering conditions ($\kappa$):} 
    As illustrated in Fig.~\ref{fig:ricean}, lower Ricean $\kappa$-factors, representing richer scattering environments, challenge more all considered localization schemes. It is shown that our approach remains effective even in highly cluttered systems with small Ricean $\kappa$-factors, consistently outperforming all benchmarks by a substantial margin. Interestingly, fingerprinting schemes appear to be greatly affected by $\kappa$.
    \item \textbf{Performance across noise regimes ($n(t)$):} Figure~\ref{fig:system_noise} showcases the superior robustness of our scheme versus noise in comparison to all considered benchmarks.
\end{itemize}
All in all, the proposed DA active sensing scheme consistently achieves superior localization accuracy than all benchmarks, while strictly adhering to the cumulative UE power budget constraint $B_P$, as expressed in the $\mathcal{OP}$ formulation.

\begin{figure}
    \centering
    \includegraphics[width=0.75\linewidth]{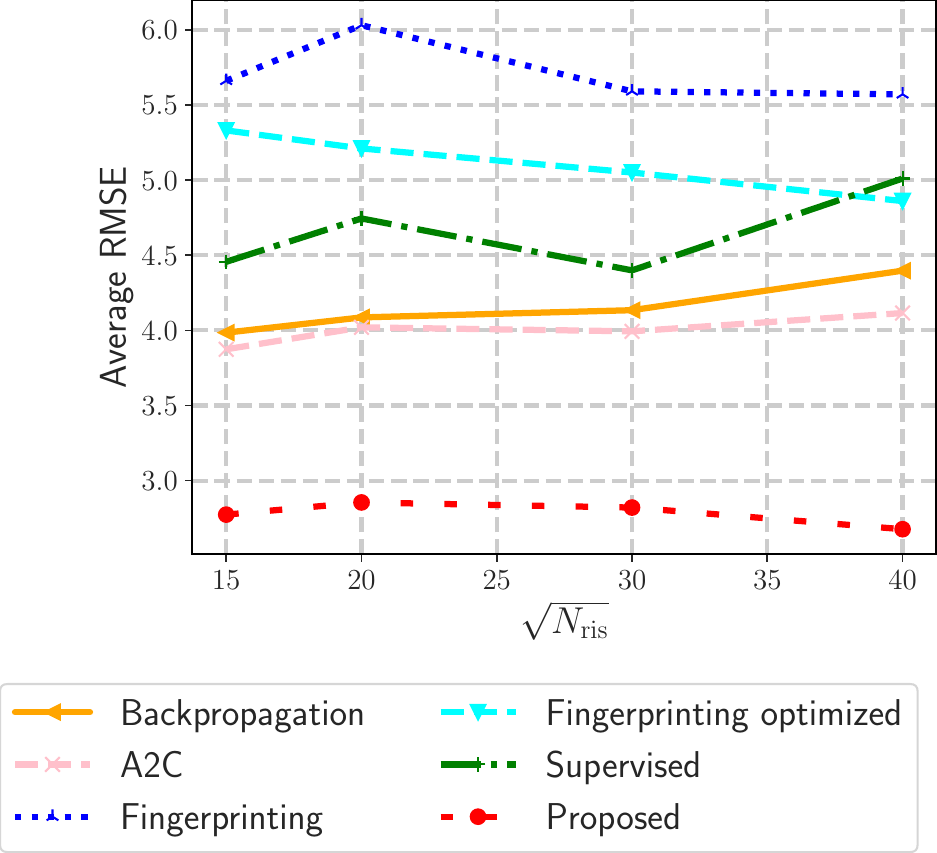}
    \caption{Localization RMSE performance versus the RIS panel size $N_{\rm ris}$, considering RSS inputs. The Ricean factor was set to $\kappa=10$ dB and the power of the noise level $\sigma_{n}$ to $-100$~dBm.}
    \label{fig:altInputSize}
\end{figure}

In Fig.~\ref{fig:altInputSize}, we evaluate the robustness of the proposed framework by repeating the RIS size scaling experiments of Fig.~\ref{fig:ris_size} using RSSs (i.e., $|y(1)|^2,\ldots,|y(T)|^2$) instead of the complex-valued received signals in baseband. The results demonstrate that our DA algorithm consistently maintains its superior performance over all baseline methods. This confirms that the learned collaborative protocol between our BS and UE agents is highly effective at extracting spatial features, even from simplified power-based observations, reinforcing our scheme's applicability to various receiver hardware constraints.

\subsection*{Pilot Transmission Power Adjustments}
\begin{figure}
    \centering
    \includegraphics[width=0.75\linewidth]{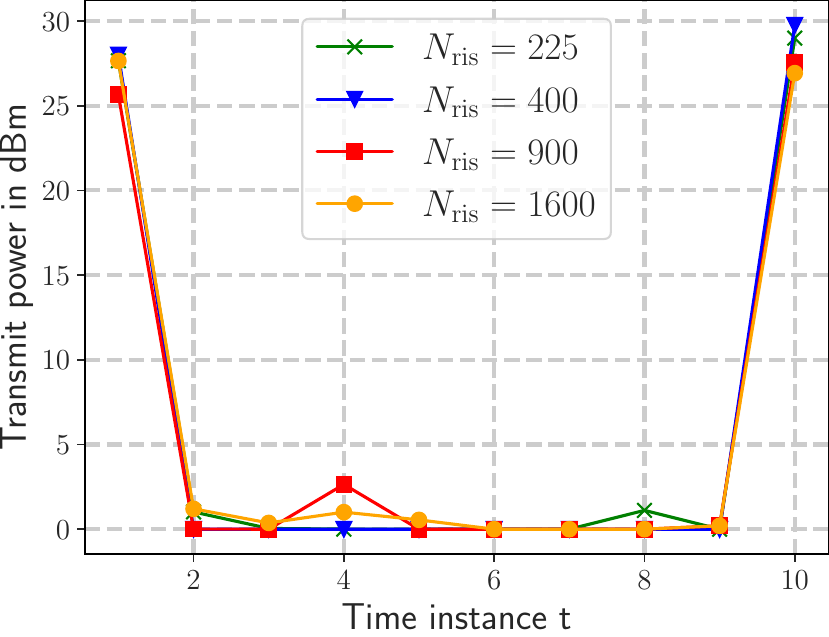}
    \caption{The average UE power levels in the uplink of the proposed active sensing scheme at each time instance $t$ for different RIS panel sizes $N_{\rm ris}$. The Ricean factor was set to $\kappa=10$ dB and the power of the noise  level $\sigma_{n}$ to $-100$ dBm.}
    \label{fig:power_level_ris_size}
\end{figure}

\begin{figure}
    \centering
    \includegraphics[width=0.9\linewidth]{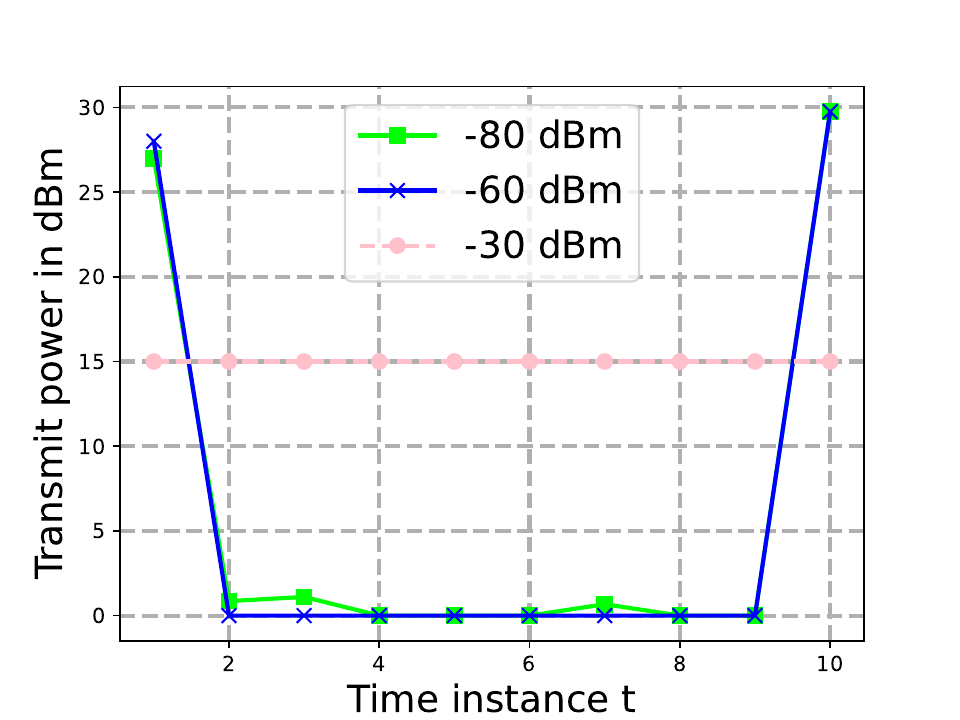}
    \caption{Same as in Fig.~\ref{fig:power_level_ris_size}, but for different  observation variance levels ($\sigma_{{n}}$). The Ricean factor was set to $\kappa=10$ dB and the RIS panel size to $N_{\rm ris}=400$. }
    \label{fig:power_level_noise}
\end{figure}

Figure~\ref{fig:power_level_ris_size} illustrates the average UE transmission power allocated at each time instance in the range $t\in [1,10]$ for the simulation setting of Fig.~\ref{fig:ris_size}. To capture the consistent behavior of the learned policy, instantaneous power levels across $1000$ independent evaluation episodes have been used for computing the average power values at each time instance $t$. It is shown that the proposed  power NN learns an interesting strategy: at the first time instance, it transmits a message with large power and then, in subsequent transmissions, the power levels become very small, often near $0$~dBm. Only in the final instance of $t=10$, a very large power level is applied to ensure detection. In Fig.~\ref{fig:power_level_noise}, we examine the effect that noise $n(t)$ has on power allocations, considering the RIS size of $N_{\rm ris}=400$. It is demonstrated that, for moderate noise levels, the two-pulse allocation of Fig.~\ref{fig:power_level_ris_size} remains consistent. On the other hand, for very strong noise, the policy learns uniform power allocation. 

\subsection*{Sensitivity Study}
\begin{figure}
    \centering
    \includegraphics[width=0.85\linewidth]{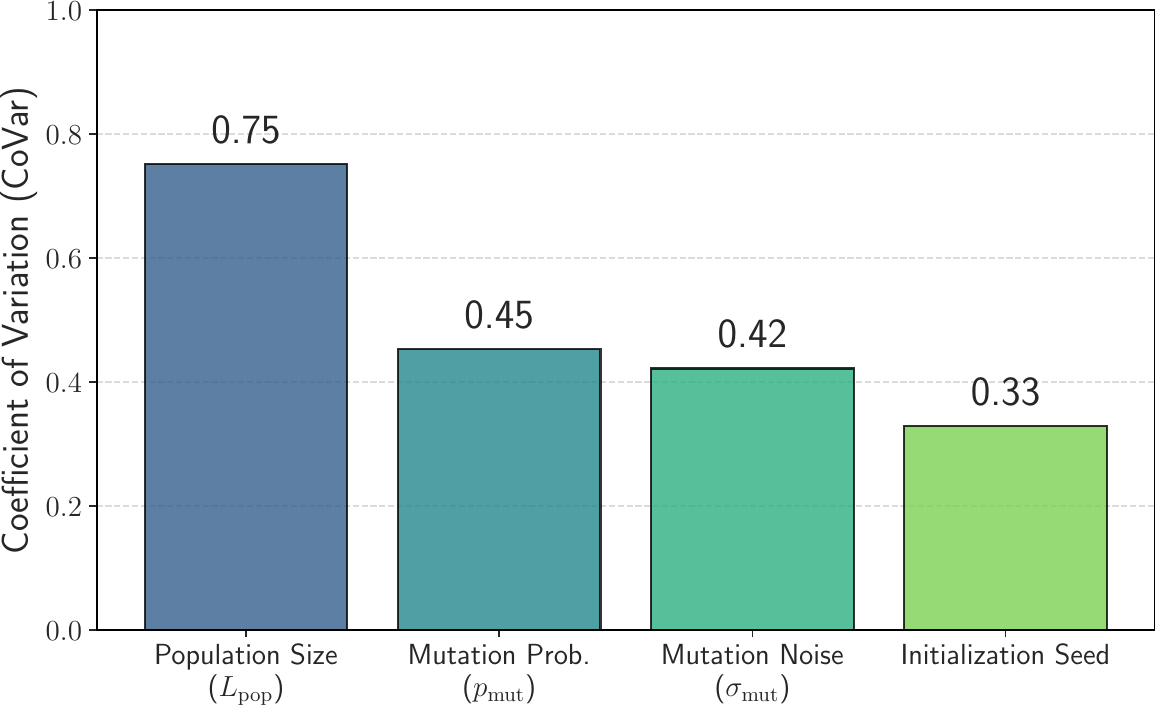}
    \caption{Sensitivity study results for tracking with $T=20$, $\sigma_{\rm ue}=0.4$, and $\omega_{\max}=0$.}
    \label{fig:sensitivity_tracking}
\end{figure}

\begin{figure}
    \centering
    \includegraphics[width=0.85\linewidth]{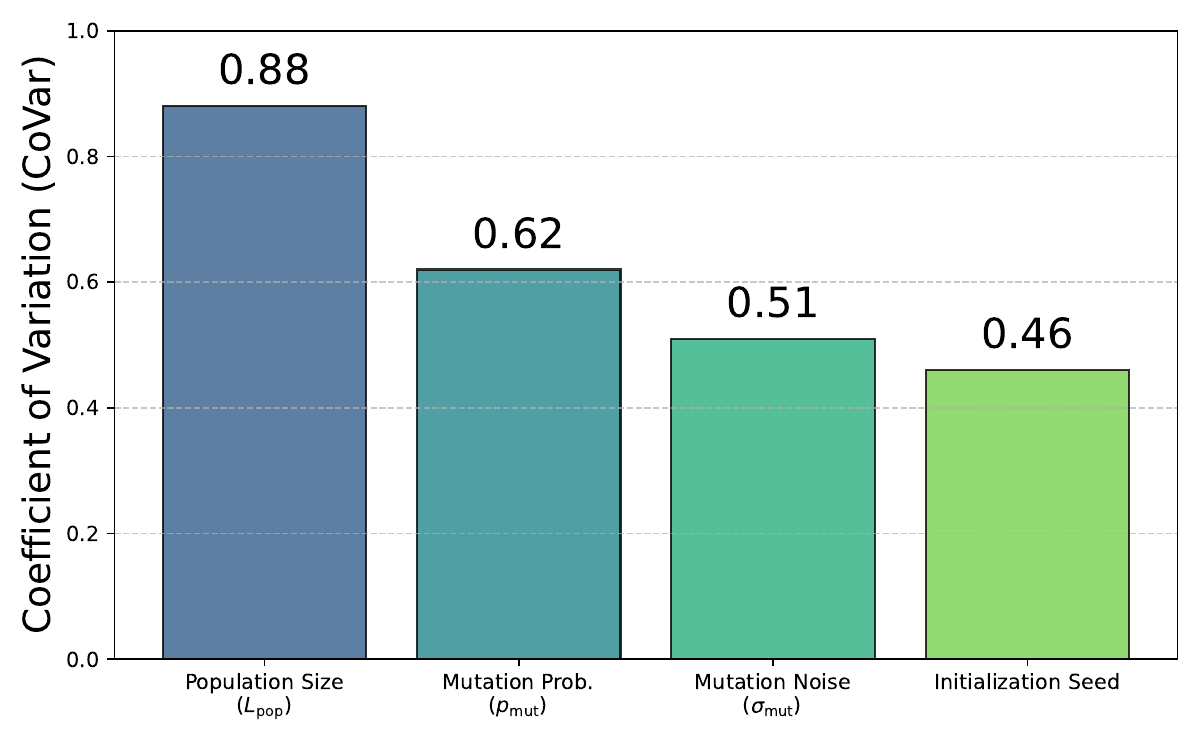}
    \caption{Sensitivity study results for localization with noise  level $n_t$ of $-60$ dBm, RIS size $N_{\rm ris}=400$, and $\kappa=10$ dB.}
    \label{fig:sensitivity}
\end{figure}

To quantify the stability of the proposed NE framework for training, we have evaluated the Coefficient of Variation (CoVar), defined as the ratio of the standard deviation to the mean. This dimensionless metric provides a normalized measure of dispersion, allowing for a direct comparison of sensitivity across hyperparameters with different scales~\cite{eaht}. In the context of stochastic optimization, a CoVar value of less than $1.0$ is considered a very good result, indicating small variance or high consistency in the algorithm's performance, since the variance is less than the mean. In our case, we have examined the sensitivity to the population size ($L_{\rm pop}$ varied from $25$ to $60$), mutation probability ($p_{\rm mut}$ varied from $0.25$ to $0.75$) and standard deviation ($\sigma_{\rm mut}$ varied from $0.25$ to $0.75)$, as well as the initialization seed ($20$ different seeds). The CoVars for a tracking and a localization system are reported in Figs.~\ref{fig:sensitivity_tracking} and~\ref{fig:sensitivity}, respectively. As depicted, these coefficients are always less than $1.0$ indicating that the proposed NE-based optimization is robust. The initialization seed produced the lowest CoVar, confirming that our three-stage training procedure, and specifically the use of an initial estimator NN $\mathcal{E}_{I}$ to guide the evolutionary process, yields a consistent and reproducible learning process.

\subsection{Impact of the Power Control Link}
\begin{figure}
    \centering
    \includegraphics[width=0.75\linewidth]{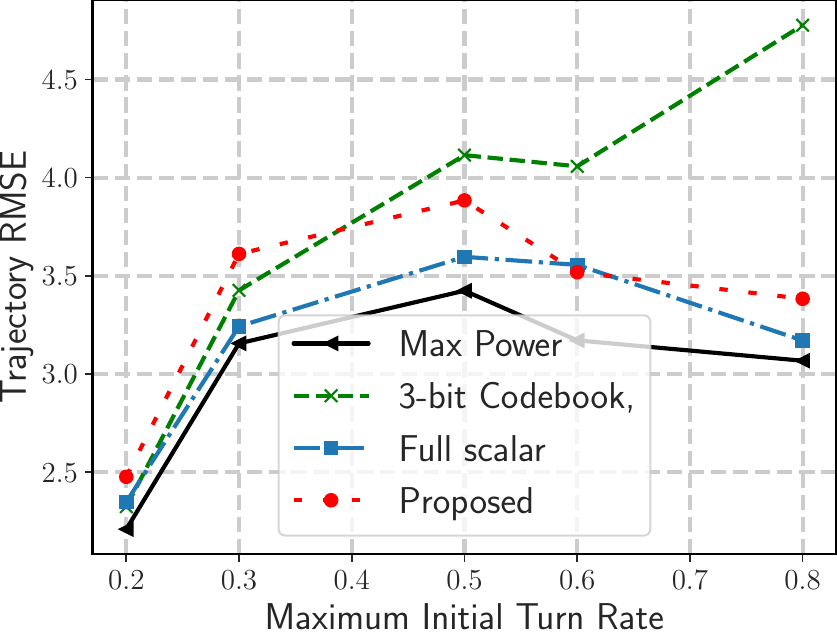}
    \caption{Same as in Fig.~\ref{fig:trackangle}, but for different control link configurations.}
    \label{fig:control_link_tracking}
\end{figure}

\begin{figure}
    \centering
    \includegraphics[width=0.75\linewidth]{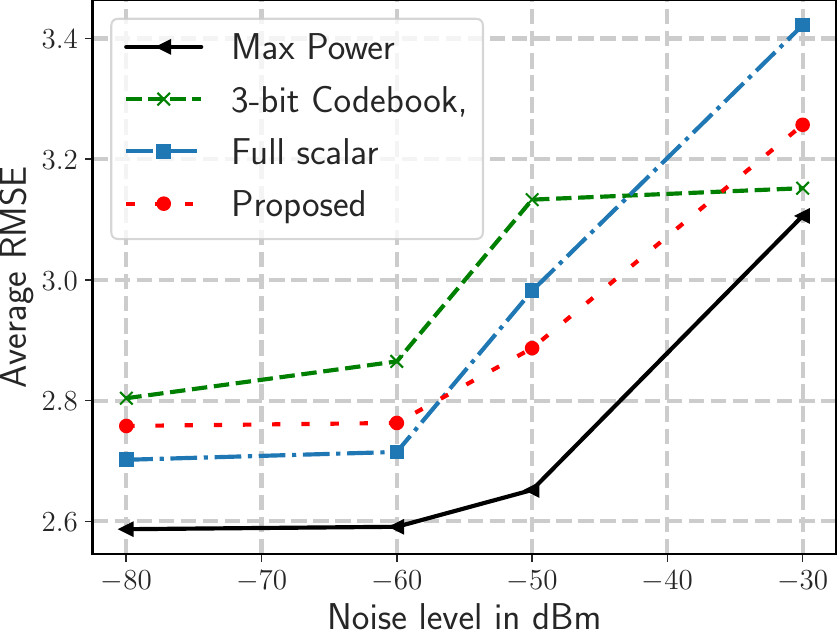}
    \caption{Same as in Fig.~\ref{fig:system_noise}, but for different control link configurations.}
    \label{fig:control_link_noise}
\end{figure}

Having established the effectiveness of the proposed NE training scheme over conventional and learning-based benchmarks, we now finally investigate the impact of the BS-UE control link configuration on the estimation performance. In Fig.~\ref{fig:control_link_tracking}, we repeated the experiment of Fig.~\ref{fig:trackangle}, with the difference that, in this case, our scheme is benchmarked against variations of NE-based algorithms. More specifically, we considered the following three NE single-agent methods: 
\begin{itemize}
    \item \textbf{Max Power:} The BS is only in charge of selecting $\boldsymbol{\Phi}(t)$; no power NN, $\mathcal{M}$, is available.
    \item \textbf{Full scalar transmission: } The BS's power NN decides on $P(t+1)$ and, then, its exact value s fed back to the UE.
     While a full scalar link allows the BS to explicitly dictate transmission power levels at the UE side, this option imposes a heavier control communication overhead.
    \item \textbf{$3$-bit codebook:} The power level $P(t)$ is encoded using a predefined $3$-bit lookup table and transmitted to the UE, which also possesses the same table.
\end{itemize}
The same NE algorithm with the same parameters was employed by all four power control schemes. Similarly, Fig.~\ref{fig:control_link_noise} performs the same comparison for the localization scenario considered in Fig.~\ref{fig:system_noise}.

Remarkably, the results in Figs.~\ref{fig:control_link_tracking} and~\ref{fig:control_link_noise} demonstrate that our DA active sensing scheme achieves tracking as well as localization performance nearly identical to the scalar and multi-bit variants across the entire noise regime. This behavior confirms that, by interpreting the temporal history of binary commands, the UE agent can effectively compensate for the information bottleneck of the considered $1$-bit power control link. Furthermore, it can be observed that the performance gap between our power-constrained  agents and the ``Max Power'' baseline remains less than $15\%$ in all test cases, despite consuming only half of the UE uplink transmission power.

\color{black}
\section{Conclusions and Future Work}\label{sec:conclusion}
In this paper, we demonstrated that active sensing can effectively enhance localization and tracking in systems with power-limited UEs and realistic RIS phase profile characteristics. A novel hybrid scheme integrating NE and supervised learning was proposed, which was shown to outperform fingerprinting baselines, constrained DRL policies, as well as backpropagation-based approaches in static localization across different RIS sizes, fading conditions, and noise levels. In dynamic tracking experiments, it was demonstrated that the proposed DA active sensing framework outperforms popular recursive state estimators with random RIS policies in a diverse set of motion models. Notably, the proposed scheme achieves high localization accuracy with only single-bit feedback power control messages, demonstrating that the learned collaborative protocol can effectively overcome information bottlenecks with negligible performance degradation compared to high capacity control links.

Future work will extend the presents framework to multi-RIS deployments and explore location privacy guarantees. We also plan to integrate advanced DNN pruning schemes \cite{stamatelisPruning} to further reduce the computational overhead for lightweight IoT devices. Finally, combining our active measurement control protocol with model-based NN tracking architectures, such as~\cite{KalmanNet,MJFNEet} is a compelling direction for future investigation.

\bibliographystyle{ieeetr}
\bibliography{references}
\end{document}